\documentclass[final,3p]{elsarticle}

\usepackage{lineno,hyperref}
\modulolinenumbers[5]

\journal{Journal of \LaTeX\ Templates}

\usepackage{graphicx}
\usepackage{amssymb}
\usepackage{epstopdf}
\usepackage{epsfig}
\usepackage{here}
\newcommand{\dd}{\,\mathrm{d}}
\begin{document}

\begin{frontmatter}
\title{Exploiting the geomagnetic distortion of the inclined atmospheric showers}

\author{Pierre Billoir}

\author{Mariangela Settimo}

\author{Miguel Blanco}

\address{%
Laboratoire de Physique Nucl\'eaire et de Hautes Energies (LPNHE), Universit\'es Paris 6 et Paris 7,\\ CNRS-IN2P3, Paris, France
}%

\begin{abstract}
  We propose a novel approach for the determination of the nature of ultra-high energy cosmic rays by exploiting the geomagnetic deviation of muons in nearly horizontal showers. 
  The distribution of the muons at ground level is well described by a simple parametrization providing a few {\em shape parameters} tightly correlated to $X^\mu_\mathrm{max}$, the depth of maximal muon production, which is a mass indicator tightly correlated to the usual parameter $X_\mathrm{max}$, the depth of maximal development of the shower. We show that some constraints can be set on the predictions of  hadronic models, especially by combining the geomagnetic distortion with standard measurement of the longitudinal profile. We discuss the precision needed to obtain significant results, and we propose a schematic layout of a detector.

\end{abstract}

\end{frontmatter}

\section{Introduction}

The nature of the UHECR (cosmic rays of ultra-high energy, of the order of 10$^{17}$ eV or more) is one of the most challenging and important open questions in astrophysics, as it is crucial for the understanding of their origin and of the mechanism of their production. The UHECR are observed through the detection of the cascade of particles they induce in the atmosphere: the first interactions occur at energies which cannot be reached in colliders, so they are described by models extrapolated above the domain where they can be fitted to the data.
As a result the mass composition of UHECR is still ambiguous and is motivating considerable efforts to improve the observation and the combination of various {\em mass indicators} and to reduce the model dependent uncertainties. Several experimental results have been obtained so far based on the observation of the shower development (e.g., depth of the shower maximum, depth of muon production), the particle content and the ground (see for example some recent publications~\cite{XmaxAuger,XmaxHiRes,XmaxTA,XmaxYakusk,lnAAuger,MPDAuger,NmuAuger}). In this paper we propose a novel method for the determination of the mass composition and the hadronic interaction models by exploring the features of horizontal air showers, impacting the atmosphere with a zenith angle larger than about 60$^\circ$. \\

One key point to identify the primary particle is evaluate the muonic component of the atmospheric shower. It is negligible within the core, and has to be evaluated in ground detectors at remote distance from the shower axis. At moderate zenith angle, the incident flux on the ground is essentially a mixture of photons, electrons, positrons (the {\em electromagnetic} component) and muons, and it is difficult to separate unambiguously the muonic components. At large zenith angle, the slant depth of the atmosphere is large, so the electromagnetic component is extinguished at ground level.  The aim of this paper is to show how the observation of the geomagnetic deviation of muons in nearly horizontal showers may help to reduce these uncertainties and provide useful cross-checks between different measurements of mass indicators.\\

The paper is organized as follows: In Sect. \ref{principle}, we describe the atmospheric shower and we explain the principle of the method; the approach used to simulate the muonic flux at ground is explained in Sect.~\ref{simulation} and we introduce in Sect.~\ref{parametrization} a convenient parameterization of the generated maps. In Sect.~\ref{correl_prim}, we show that the coefficients of this parametrization are tightly correlated to the nature of the primary particle and we examine their dependence on the models of hadronic interactions; in Sect. \ref{precis_meas}, we evaluate the precision needed on ground measurements to obtain predictive results and, in Sect.~\ref{detector}, we propose possible layouts for a detector.

\section{Principle of the method}\label{principle}

\par Extensive atmospheric showers produced by protons or nuclei  include a {\em hadronic} component (mainly charged mesons) and an {\em electromagnetic} component (photons, electrons and positrons) induced by the decay of neutral mesons (mainly pions) into photons. These components reach their maximal development after crossing between 500 and 1000 g/cm$^2$, depending on the nature and the energy of the primary particle, and then decrease progressively. The depth of maximal size $X_{\rm max}$ (where the number of charged particles, mainly electrons and positrons, is maximal) is known and exploited for a long time, and may be considered as a standard; it is essentially sensitive to the electromagnetic component.

\par For moderately inclined showers, both components reach ground level, and their relative importance (especially the muonic content) is in principle an indicator of the nature of the primary, but the techniques needed to evaluate separately the components are not straightforward. At large zenith angles $\theta$ (typically $\theta > 60$ deg) the electronic cascade is extinguished and the hadronic one has been transformed through meson decays into a flux of muons which do not interact strongly. These muons lose their energy mainly through ionization, and many of them decay in flight, but a significant fraction reaches the ground. Their initial azimuthal distribution is uniform around the shower axis;  but this symmetry is broken by the deviation in the magnetic field of the Earth, which results in a distortion of the density at ground level. A detailed discussion of this effect may be found in \cite{param_hor_1}. \\

\par
\begin{figure}[H]
\begin{center}
\epsfig{file=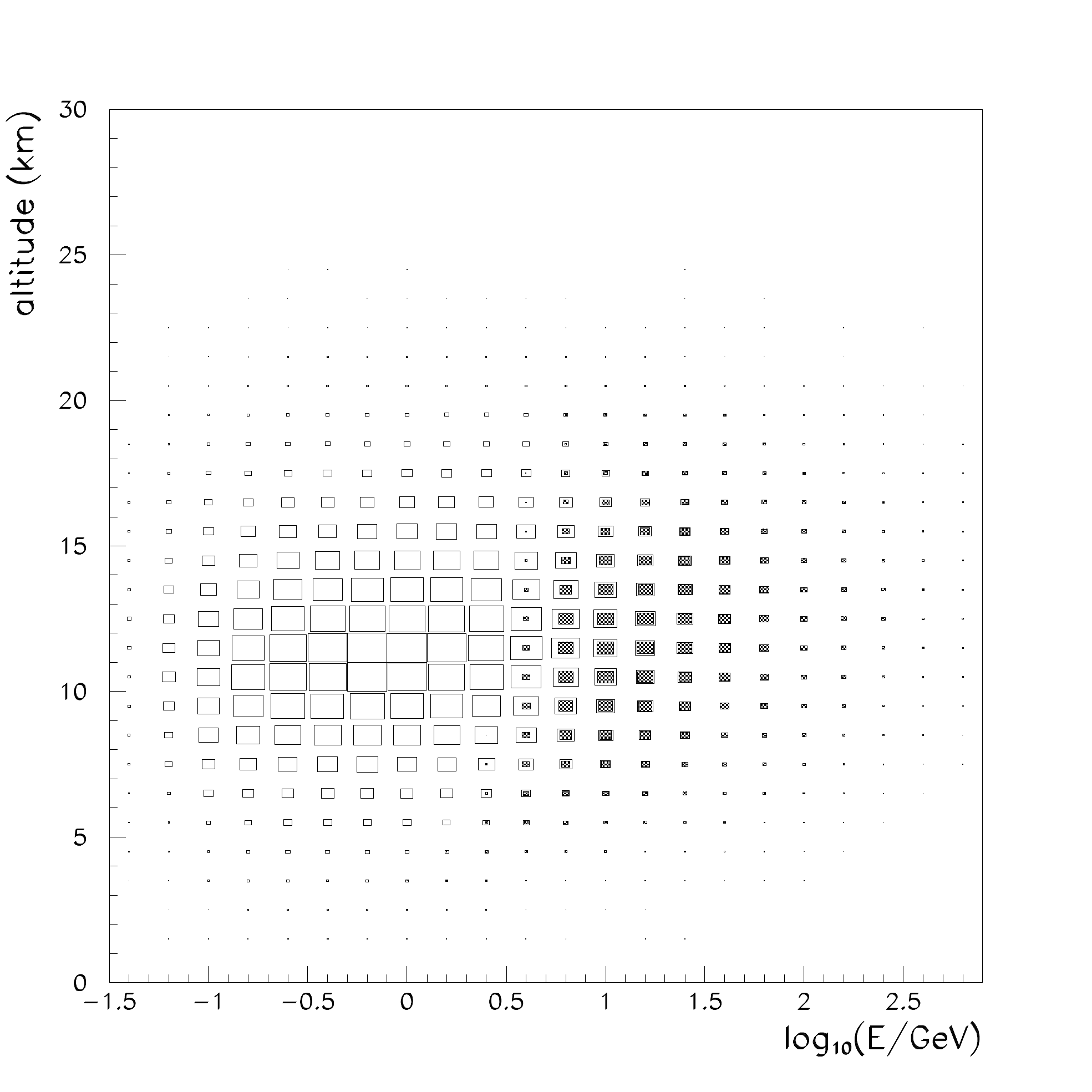,width=6.5cm}
\epsfig{file=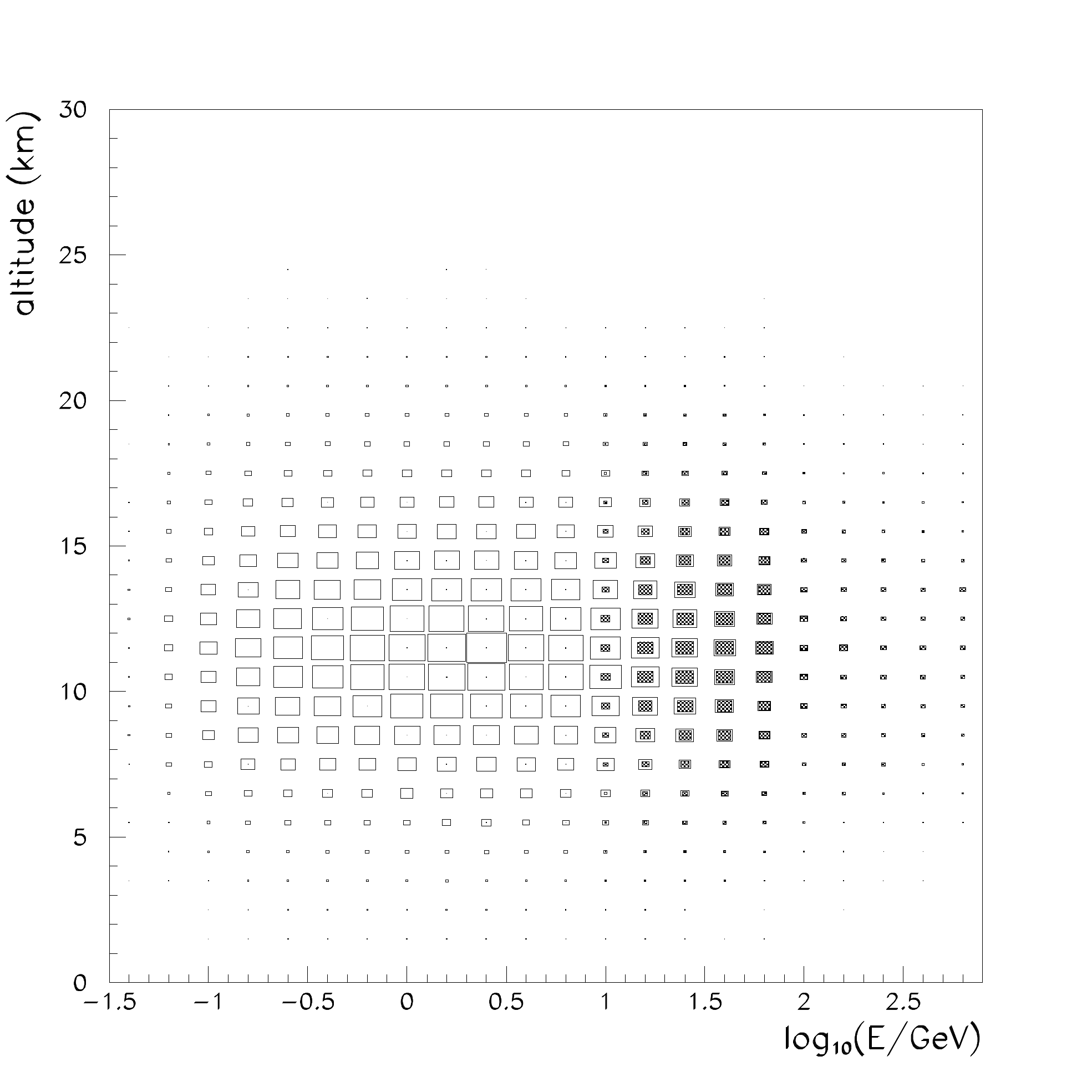,width=6.5cm}
\caption{\footnotesize Distribution in altitude and energy of the muons
  produced in inclined showers; in grey: muons reaching the
  ground. Left: proton shower of 1 EeV, $\theta=72$ deg; right: 80 deg.}
\label{fig:prod_altit_ener}
\end{center}
\end{figure}

 The distribution of the muons in altitude and energy is illustrated in Fig. \ref{fig:prod_altit_ener}: for showers with a zenith angle around 75 deg, the energy of most muons reaching the ground is of the order of a few GeV to a few 10 GeV, and their path is a few 10 km, so the magnetic deviation is of the order of a few 100 m: the distortion is sizeable and may be measured by a surface detector with a sufficient granularity. 
\par
\begin{figure}[H]
\begin{center}
\epsfig{file=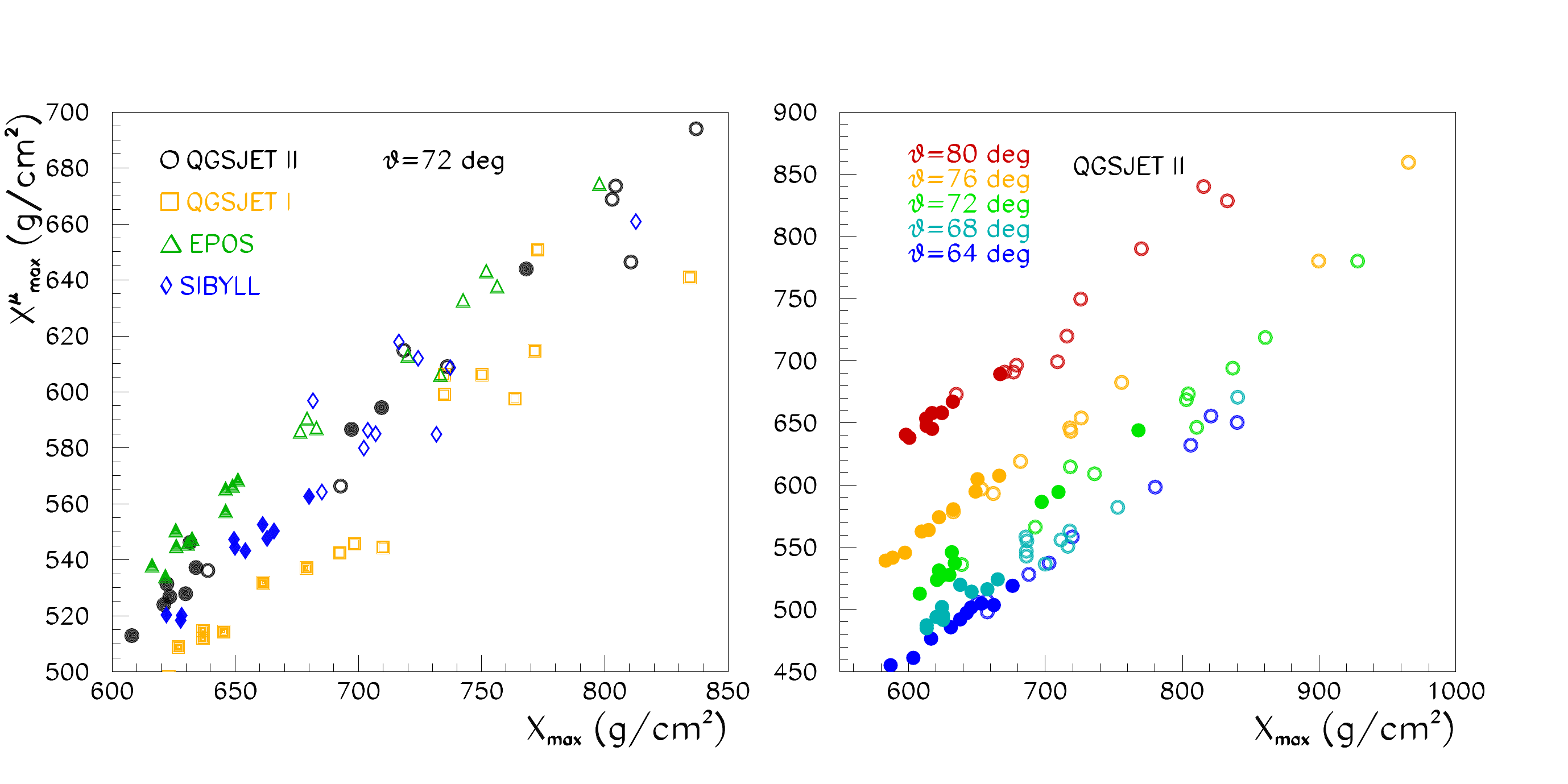,width=15cm}
\caption{\footnotesize Correlation between the depth of maximal size $X_{\rm max}$ and the depth o maximal muon production $X^\mu_{\rm max}$.
Left: Showers at $\theta=72$ deg, from protons (open symbols) or iron nuclei (solid symbols) using different models for hadronic
interactions (see below Sect. \ref{dep_model}. Right: dependence on zenith angle for the QGSJET II model.}
\label{fig:xmax_xmumax}
\end{center}
\end{figure}

 The deviation is proportional to the square of the path of the muons down to the ground and to the inverse of their energy,  which increases in average with the path because of the losses through decay. The net effect is an increase of the distortion with $\theta$, and we can expect the distortion to depend also on the longitudinal evolution of the cascade, and especially on $X^{\mu}_{\rm max}$, the depth of maximal production of muons, which is tightly correlated to the standard parameter $X_{\rm max}$ for a given value of the zenith angle $\theta$, as it is shown in Fig. \ref{fig:xmax_xmumax}.
\par The dependence on $\theta$ is due to a density effect: for higher $\theta$, $X_{\rm max}$ is reached at higher altitude, where the interaction length is larger, so the mesons decay earlier in terms of atmospheric depth. Both $X_{\rm max}$ and $X^{\mu}_{\rm max}$ are indicators of the nature of the primary particle when using a reliable model of the interactions at very high energy, especially the hadronic ones.
\par Our aim is to express this dependence through a simple paramerization of the muon density and provide tools to extract from the observation  of inclined showers an estimator of the mass composition using a given model of hadronic interactions at ultra-high energy; we want also to obtain some discrimination between different models through their prediction of the correlation between $X_{\rm max}$ and $X^{\mu}_{\rm max}$. This approach has the advantage of using the pure muonic component of the shower. Moreover in inclined showers the hadronic cascade is observed at a higher energy level than in nearly vertical showers, because  the mesons decay earlier. So the information is complementary to studies at low zenith angles. 
\par Taking average values of the parameters, we obtain a simple analytic expression of the density that can be used as an alternative
to previous propositions (\cite{param_hor_1},\cite{param_hor_2}), which did not account for the variation of the longitudinal profile.

\par
\begin{figure}[H]
\begin{center}
\includegraphics[width=12cm]{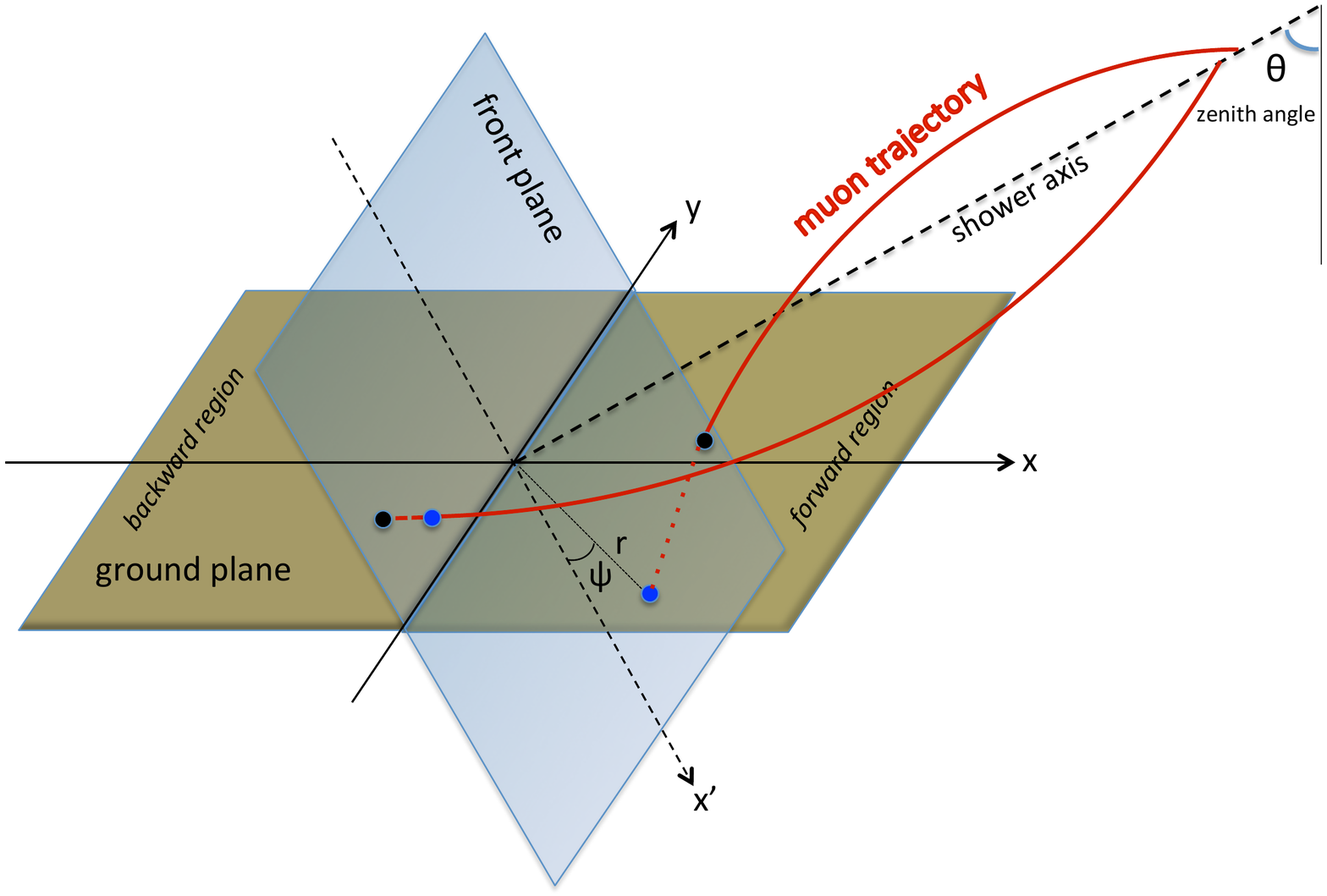}
\caption{\footnotesize Geometry of an inclined shower: the front plane
  is perpendicular to the shower axis, which makes an angle $\theta$
  with the vertical direction. A point in the front plane is defined by the polar coordinates $r,\psi$.}
\label{fig:geometry}
\end{center}
\end{figure}

 The geometrical frames and coordinates used hereafter are described in Fig. \ref{fig:geometry}. 

\section{Simulation of the muonic component}\label{simulation}

\subsection{General procedure}

\par In this paper we estimate the muon density through a procedure in two stages, to reduce the computing load:
\begin{itemize}
\item Extract from a shower simulation package the list of the muons at their production point. We simulate proton and iron showers at different primary energies between $10^{17}$ and $10^{19}$ eV,  and zenith angles between 64 and 80 deg.
\item From each sample, propagate the muons to obtain the density on the ground or in a {\em front plane} (orthogonal to the shower axis, see Fig. \ref{fig:geometry}), with different values of the transverse magnetic field, which is the only relevant quantity because the muons that reach the ground are nearly parallel to the axis. 
\end{itemize}

We do not try to describe the core of the shower (distance less than about 100 m), nor the behaviour at large distances ($\gg$ 1 km), where the density is well below one muon per square meter, so that detectors of a reasonable size have little chance to make a precise measurement. We suppose that the detectors cannot distinguish the charge of the muons, so we consider only the total flux as measurable.\\

\par To transpose the geometrical observations into a description in terms of the depth $X$, the vertical profile of the atmosphere as a function of the altitude $X(h)$ needs to be known at any time, not only to define the stage of evolution of the hadronic cascade, but also to describe properly the rate of decay of mesons, which depends on the density $\dd X/\dd h$ . Fortunately, this profile has little diurnal and seasonal variations at altitudes between 10 and 30 km above sea level \cite{atm_keilhauer}, where most of the muons are produced in the showers  we are interested to. So in the simulation of the showers we use the standard Linsley atmosphere \cite{linsley}. For the propagation of the muons we use either the same model, or a pure exponential profile to obtain analytical evaluations. 
\\
\par Through their decays and their radiative interactions initiating electromagnetic cascades of low energy, the muons generates continuously the so-called ``electromagnetic halo'' (see for example \cite{halo}).  These cascades develop over a short distance compared to the muon path to the ground, so the electromagnetic flux follows closely the shape of the muonic one, and undergoes the same magnetic distortion. The impact on surface measurements (essentially a constant factor in most cases) depend on the nature of the detectors, and is beyond the scope of this paper. So we do not simulate the electromagnetic halo.

\par The energy loss $\varepsilon=-\mathrm{d}E/\mathrm{d}X$ is supposed to be constant (2 MeV.g$^{-1}$cm$^2$). This is a good approximation in the medium range of energy (few 100 MeV to 10 GeV). Very energetic muons lose more per unit of depth, but the total loss is anyway a small fraction of their   initial energy; moreover they are concentrated in the core, and are weakly deviated. Muons of low energy stop rapidly and contribute little to the density at  ground. \\

\subsection{Simulation tools}

To simulate the atmospheric shower we use the package CORSIKA \cite{CORSIKA}. We set options for the {\em thinning} procedure such
that the statistical weight of the hadrons remains low (less than 10 within the pure hadronic cascade and less than 1000 for hadronic
subproducts of the electromagnetic cascade through a photo-nuclear interaction).
\par
\begin{figure}[H]
\begin{center}
\epsfig{file=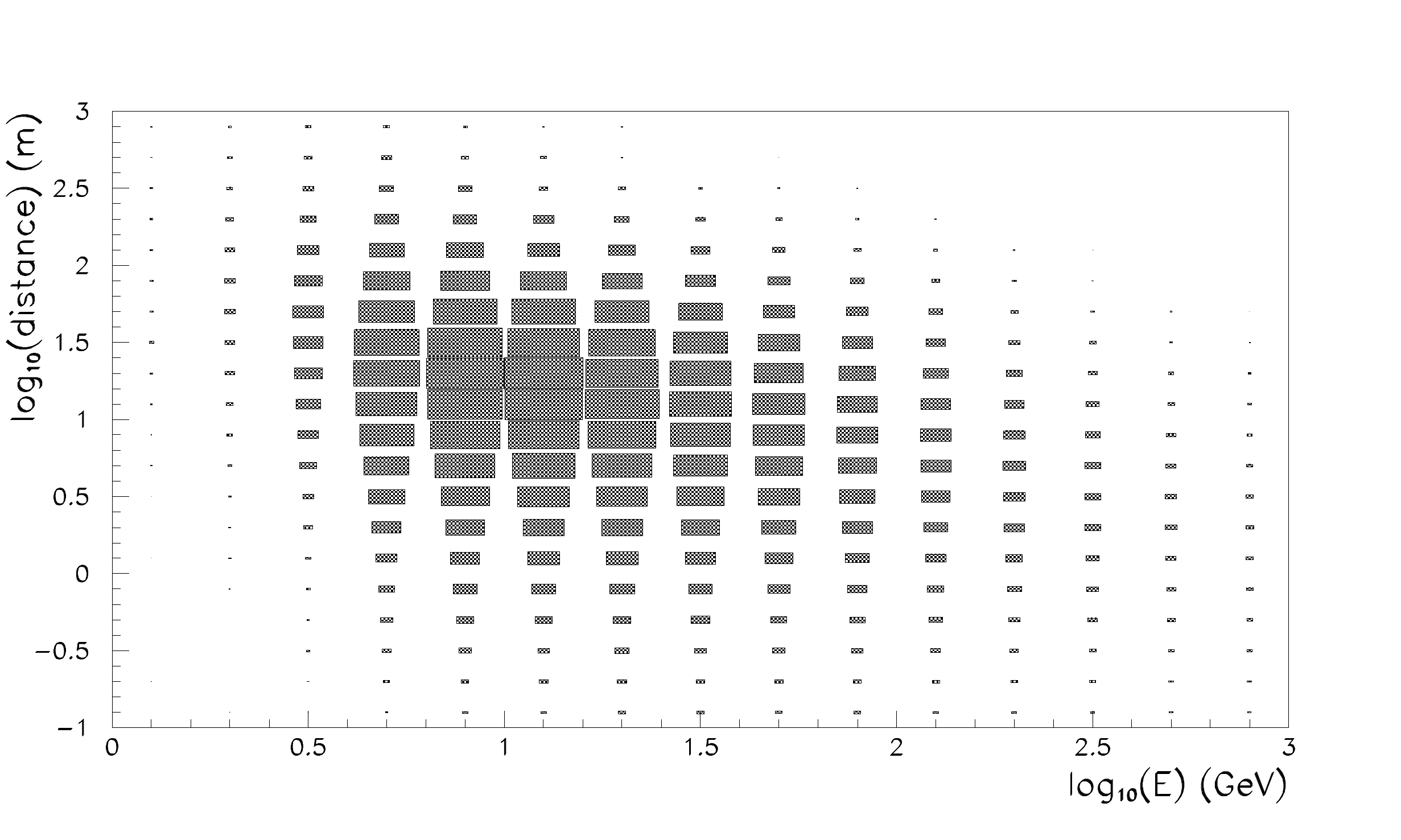,width=12cm}
\caption{\footnotesize Distribution in distance from axis vs energy at production point, for muons
  produced in proton showers of 1 EeV, $\theta=72$ deg, reaching the ground.}
\label{fig:prod_dist_ener}
\end{center}
\end{figure}

 CORSIKA offers an option to output the list of the muons generated by the hadronic cascade, with their production point, their
momentum at this point and their statistical weight $w$ inherited from the parent particle. The production region of the muons reaching the
ground is mainly concentrated within a few ten meters around the shower (see Fig. \ref{fig:prod_dist_ener}), and this distance is small
compared to both the magnetic deviation and the dispersion due to the multiple Coulomb scattering. So we keep only the longitudinal 
position of the origin of the muon.

\par The original direction of the muon is slightly affected by the magnetic deviation of its parent (most of the time a pion or a kaon),
which has the same charge and a similar energy. For a given momentum
$p$ the mean flight is
$\overline{l}=\tau p/m$; the ratio $\tau/m$ is much smaller for a kaon than for a pion, so we will discuss only the pion case.  
The angular deviation over a length $\overline{l}$ is $\delta=cB_t\overline{l}/p$ where $B_t$ is the transverse field (in Tesla) and $p$ the momentum in
eV$/c$, that is $\delta=cB_t\tau/m$; with $B_t<6.10^{-5}T$ for the field at the surface of the Earth, we see that $\delta
\lesssim 1$ mrad anywhere. This value is small compared to the deviation of the produced muon along its path to the ground, and we have
checked that changing systematically the initial directions of the
muons by $\pm$1 mrad (depending on the charge) does not modify significantly our results. So we assume a perfect
symmetry of the directions around the axis and to obtain smooth density maps we replace every muon by $N$ clones obtained by successive
rotations of $2\pi/N$ around the shower axis, each one with a weight $w/N$.\\

\par 
This sample is the input for two different procedures
aimed to obtain the map of the muon flux on the ground:
\begin{itemize}
\item a stepwise extrapolation of the muons through the atmosphere down to the ground, accounting for
 energy loss, multiple scattering, decay probability and magnetic deviation in each step. For a muon of weight $w$, $N$ is chosen as
 the integer just above $w/0.2$, so the clones have a weight less than 0.2. 
\item an analytic computation of a density of probability in the front plane for each input muon, in the approximation of an exponential atmosphere
  profile, and a weighted summation of these densities. In this option $N=24$.
\end{itemize}
These procedures are detailed in the following subsections. 

\subsection{Stepwise approximation}\label{step_extrap}

Each clone is propagated by steps (initially 1 km) until the muon stops, decays or reaches the ground. The energy
loss is computed according to the local density of air, which is defined as a function of the altitude.  The step
length is divided by 2 if the muon loses more than the half of its energy. In each step the position and the
direction undergo a deterministic deviation from the magnetic field,
and a random one to account for the multiple scattering \cite{PDG}. 
When a step goes below the ground level, a linear interpolation gives the impact at the ground.
It was checked that reducing the step length did not modify significantly the final distribution.
\par From the set of surviving muons, we want to define a ground density. A first option is just to count the number of muons per unit
of ground surface. Doing so we observe a forward-backward asymmetry due to the divergence of the muons: the muons hitting the ground in
the upstream region have a large angle of incidence. We can also define an {\em intrinsic} density as the number of muons hitting a
unit surface in a plane orthogonal to the shower axis. In that case there is still an asymmetry because the upstream and downstream impacts are
not at the same longitudinal position in the shower. In practice what matters is the number of muons entering a detector; if the detector is
a planar vertical surface, the intrinsic density is more relevant to
estimate the response. We will show in Sect. \ref{compar_step_anal} that the forward-backward asymmetry
does not prevent us from exploiting the magnetic distortion in a reliable way.

\subsection{Analytic calculations for an exponential  atmosphere}\label{analytic}

In the local ground frame ($z$ axis upwards on the vertical direction, with the ground at $z=0$),
the atmospheric depth (quantity of matter above altitude $z$) is described by :
\begin{equation}
X(z) =  X_a \exp(-z/L)
\end{equation}
where $L$ is the attenuation length (typically 8 km). We define the {\em shower frame} with a coordinate  $z'$ along the
shower axis (origin at ground level), $y'$ perpendicular to the shower axis in the horizontal plane (see Fig. \ref{fig:geometry}).
An inclined shower with a zenith angle $\theta$ is considered as equivalent to a vertical one in an
exponential atmosphere with a scaled attenuation length:
\begin{equation}
X(z') =  \frac{X_a}{\cos\theta} \exp\left(-\frac{z'}{L/\cos\theta}\right) = X_{sl} \exp\left(-\frac{z'}{L_{sl}}\right)
\end{equation}
where $X_{sl}$ is the slant depth at ground level and $L_{sl}$ the
attenuation length along the axis.
Actually there is a transverse gradient of density in the shower frame, so this expression is valid only around the axis, at a short
distance compared to $L$: we show in Sect. \ref{compar_step_anal} that this asymmetry, combined with the fact that the ground is replaced by a
plane orthogonal to the shower axis, has little consequence on the evaluation of the distortion.\\

\par Let us consider a muon (mass $m$, lifetime $\tau$) injected at $z'_i$, that is at a slant depth $X_i=X_{sl}\exp(-z'_i/L_{sl})$),
 with a kinetic energy $E_i$, nearly parallel to the shower axis. We define $\lambda= c\tau/m$, $E_\infty=E_i+\varepsilon X_i$
 (energy extrapolated backwards to infinite altitude).

In Appendix we obtain the following results:

\begin{itemize}
\item  The muon can reach the ground if $E_{gr}=E_\infty-\varepsilon X_{sl} > 0$, with a probability
\begin{equation}
P_{gr}=\left(\frac{E_{gr}X_i}{E_i  X_{sl}}\right)^{\frac{L_{sl}}{\lambda  E_\infty}}
\end{equation}
(this expression was already derived in \cite{cazon}  within the same approximations)

\item The magnetic angular deviation (perpendicular to the
  transverse field $\overrightarrow{B_t}$) is:\\
\begin{equation}
\omega = \frac{\beta}{E_\infty\cos\theta} \left( z'_i+L_{sl}\,\ln\frac{E_i}{E_{gr}}\right)
\end{equation}
(with $\beta=ecB_t$ or $\beta=cB_t$ if the energies are expressed in eV)\\
and the transverse displacement from a straight line:
\begin{equation}
\delta = \frac{\beta L_{sl}^2}{E_\infty} F_1(\alpha,z'_i/L_{sl}) 
~~~\mathrm{with}~~~
\alpha = \frac{\varepsilon X_{sl}}{E_\infty}~~,~~F_1(\alpha,\zeta) = \int_0^\zeta \frac{u\dd u}{1-\alpha\exp(-u)}
\end{equation} 

\item The variance of the angular deviation due to the multiple scattering is:
\begin{equation}
\sigma_{ang}^2 =  \frac {E_{ms}^2}{\varepsilon X_{rad}} \left(\frac{1}{E_{gr}}-\frac{1}{E_i}\right)
\end{equation}
and the variance of the displacement:
\begin{equation}
\sigma_{pos}^2 =  \left(\frac{E_{ms}}{E_\infty}\right)^2 \frac{X_aL_{sl}^2}{X_{rad}} F_2(\alpha,z'_i/L_{sl})
\end{equation}
with 
\begin{equation}
F_2(\alpha,\zeta) = \int_0^\zeta \frac{u^2\dd u}{(1-\alpha\exp(-u))^2}
\end{equation}

\end{itemize}

With these expressions we can attribute a weighted density in the ($x',y'$) plane to each muon emitted with a weight $w$ at $z'_i$ 
along the unit vector ($u'_x,u'_y,u'_z$); for example, if $\vec{B}_t$ is along $y'$ axis, so the deviation along $x'$:
 
\begin{equation}
f(x',y') = w\,P_{gr} \frac{1}{2\pi\sigma_{pos}^2}\,\exp\left(-\frac{(x'-u'_xz'_i\pm\delta)^2+(y'-u'_yz'_i)^2}{2\sigma_{pos}^2}\right)
\end{equation}
where $\pm$ represents the sign of the muon charge.

\section{Parametrization of ground densities}\label{parametrization}

\subsection{General features}

\begin{figure}[H]
\begin{center}
\epsfig{file=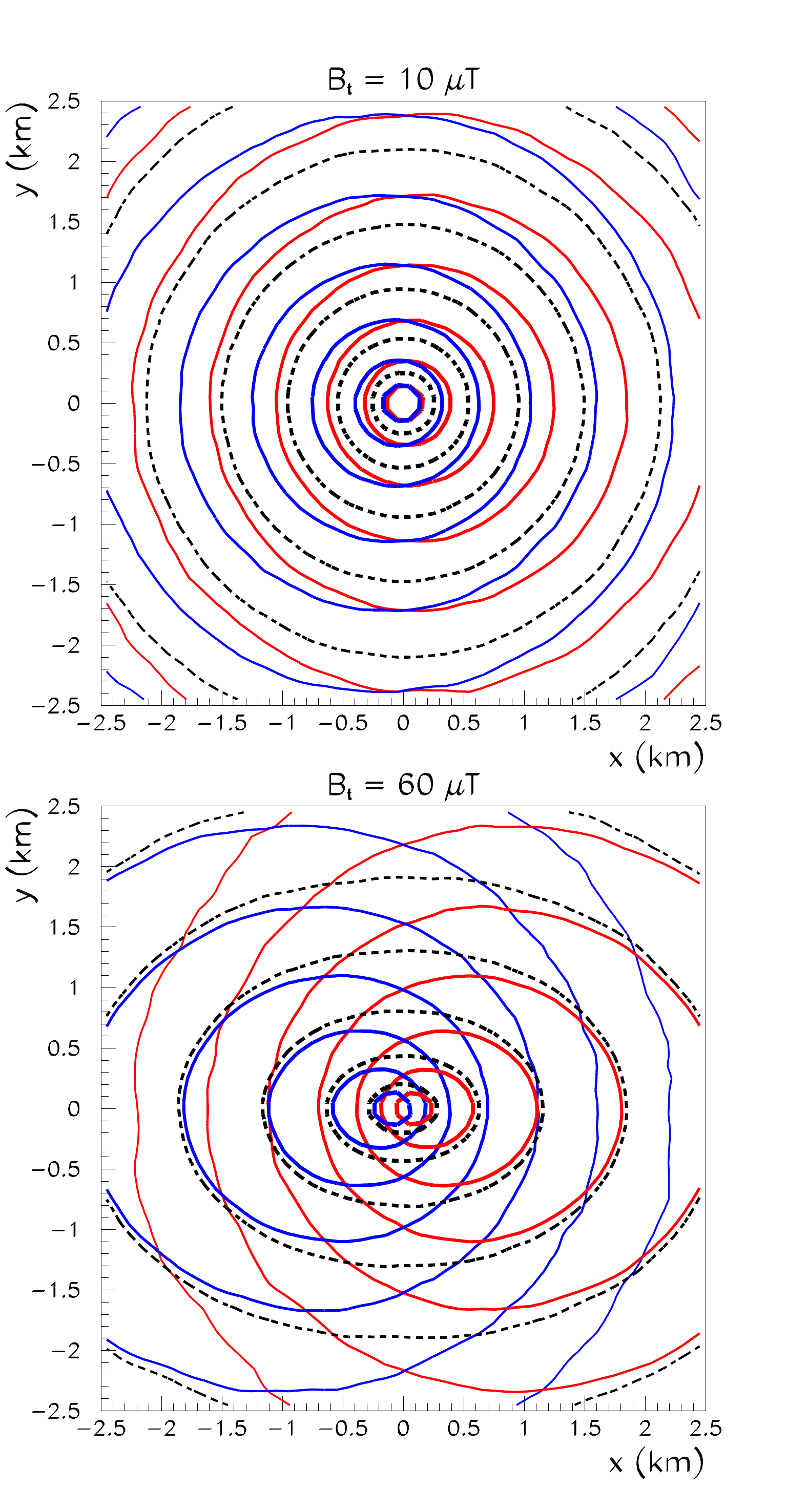,width=5cm}
\hspace{-5mm}
\epsfig{file=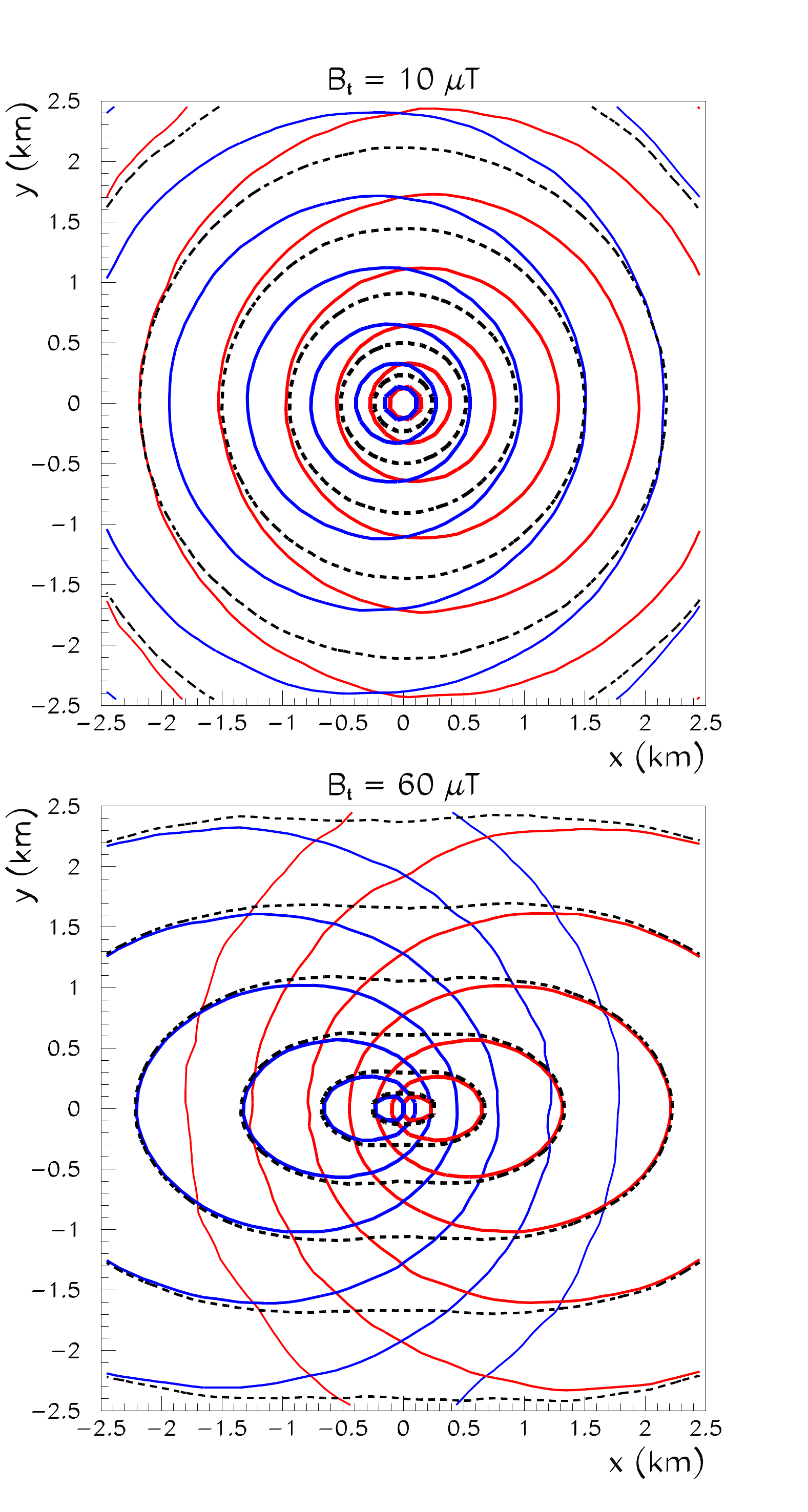,width=5cm}
\hspace{-5mm}
\epsfig{file=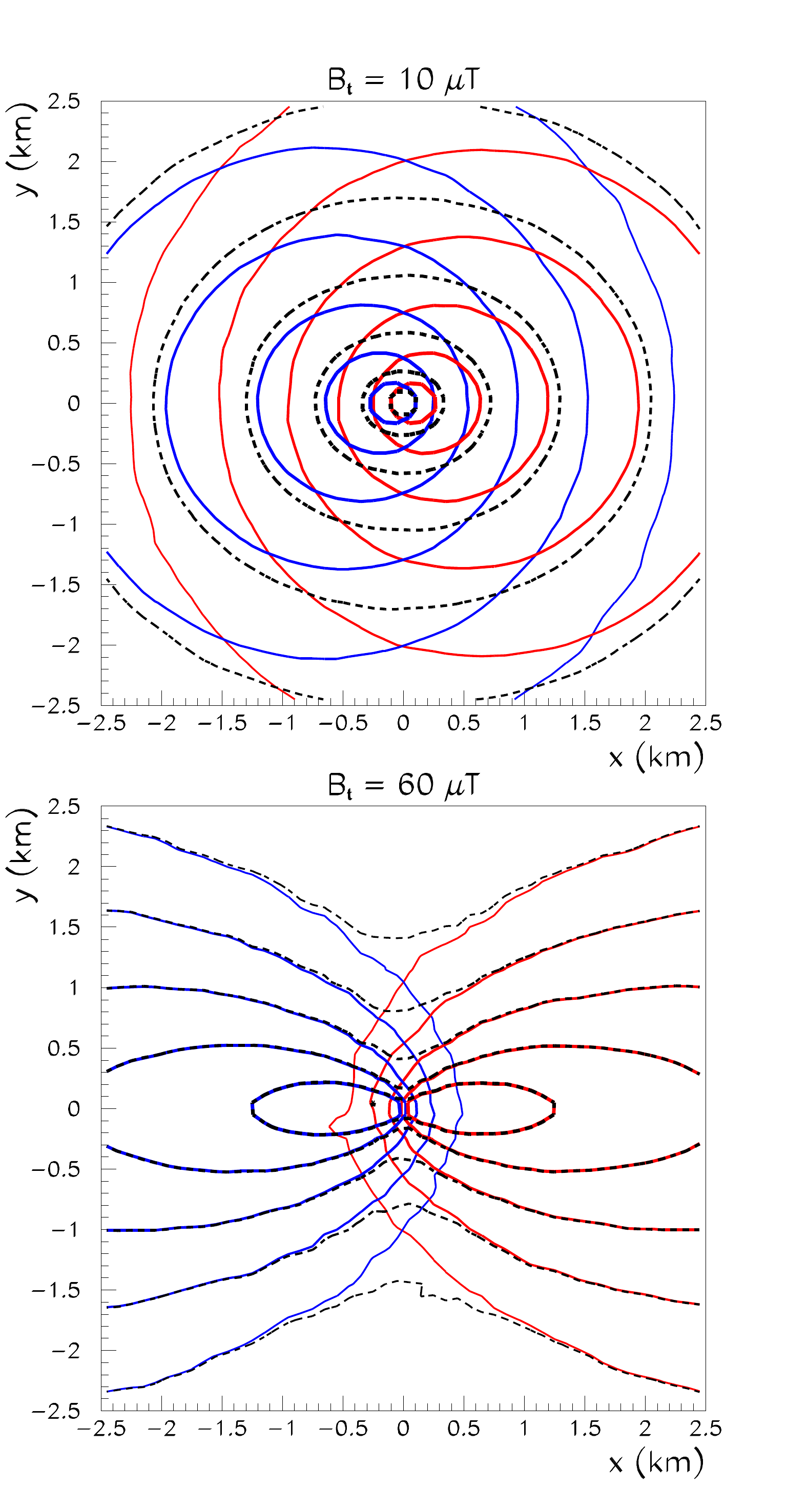,width=5cm}
\caption{\footnotesize Contour levels of the muon density in the transverse plane,
  for a proton shower of 10 EeV at 3 zenith angles (from left to right:
  64, 72 and 80 deg), with a transverse field of 10 or 60 $\mu$T along $y$ axis. In
  red: $\mu^+$, in blue:$\mu^-$, in black (dashed): total. The lines
  correspond to equidistant levels in log scale (2 per decade),
  starting from 10$^{-2}$ muons/m$^2$.}
\label{fig:map_xy}
\end{center}
\end{figure}

 As expected, the magnetic distortion increases with the zenith angle $\theta$, and with the magnitude of the transverse field
 $B_t$.  This is illustrated in Fig. \ref{fig:map_xy} which shows the contour levels of the densities of muons (total or charge separated). The
 dependence of $B_t$ is not trivial; we can distinguish two extreme cases:
\begin{itemize}
\item weak distortion (low $\theta$ and $B_t$): the total density is the sum of the positive and the negative components: at a given position,
  each one differs from the undistorted one by a small amount, which is proportional to $B_t$, with opposite signs, so the global effect
  cancels at first order, and the variation is of the second order in $B_t$.
\item strong distortion (large $\theta$ and $B_t$): almost everywhere, the flux is fully dominated by muons of one sign, so the dependence
  on $B_t$ is stronger.
\end{itemize}

\subsection{Functional parametrization}\label{function}

\begin{figure}[H]
\begin{center}
\epsfig{file=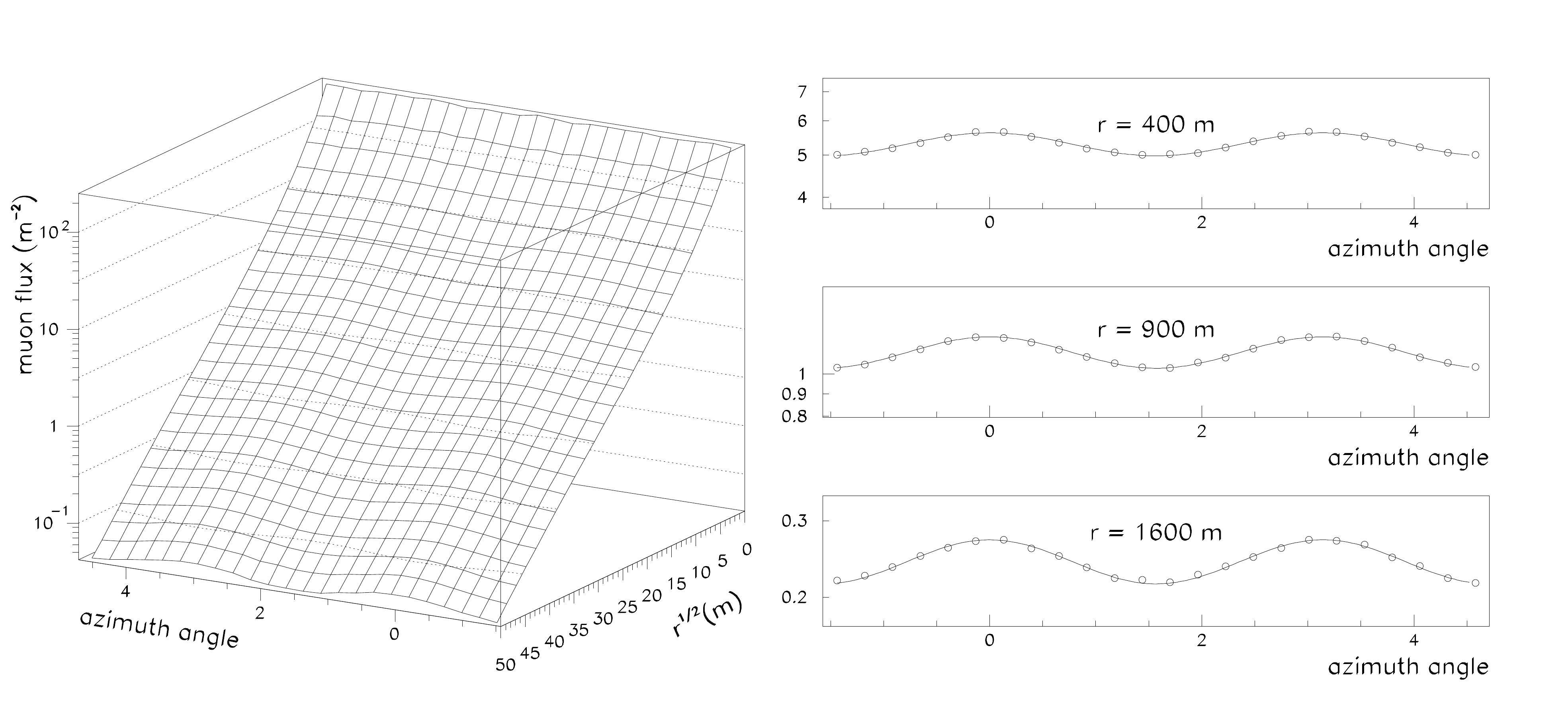,viewport=0 10 1100 440,width=14cm}
\epsfig{file=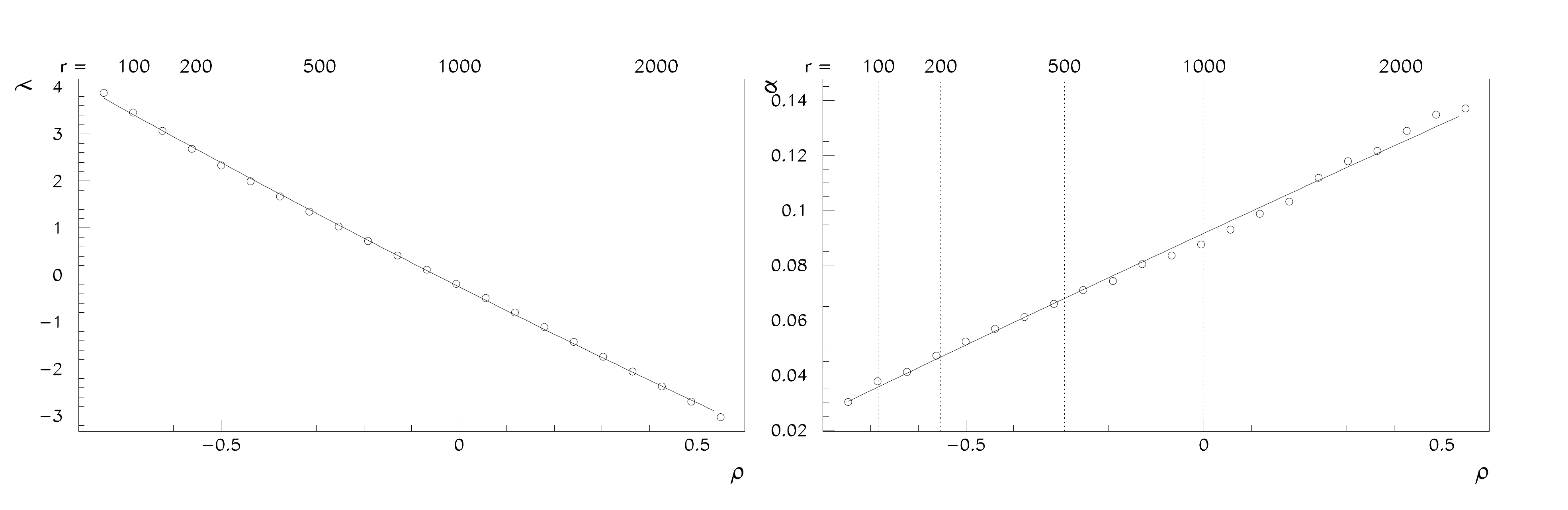,viewport=0 10 1100 340,width=14cm}
\caption{\footnotesize Parametrization of the muon density in ($r,\psi$) coordinates, for a proton shower at 10 EeV, $\theta=64$
deg, $B_t=30~\mu$T. Top left: density as a function of $\sqrt{r}$ and $\psi-\psi_B$; top right: azimuthal dependence at different distances;
bottom left:  parameter $\lambda$ (logarithm of the density) as a function of $\rho=\sqrt{r/r_\mathrm{ref}}-1$; bottom right:
parameter  $\alpha$ (azimuthal modulation) as a function of $\rho$.} 
\label{fig:map_rphi_1}
\end{center}
\end{figure}

\begin{figure}[H]
\begin{center}
\epsfig{file=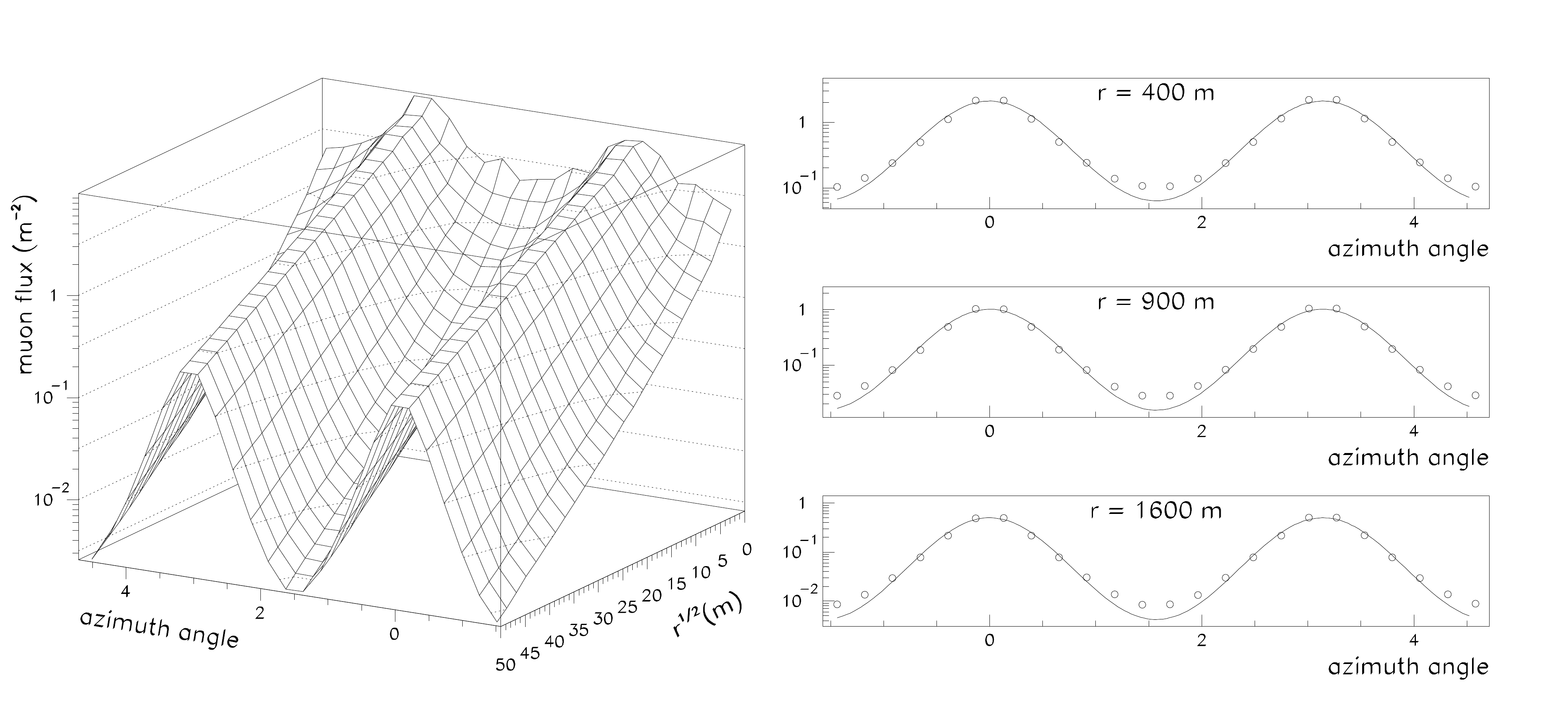,viewport=0 10 1100 440,width=14cm}
\epsfig{file=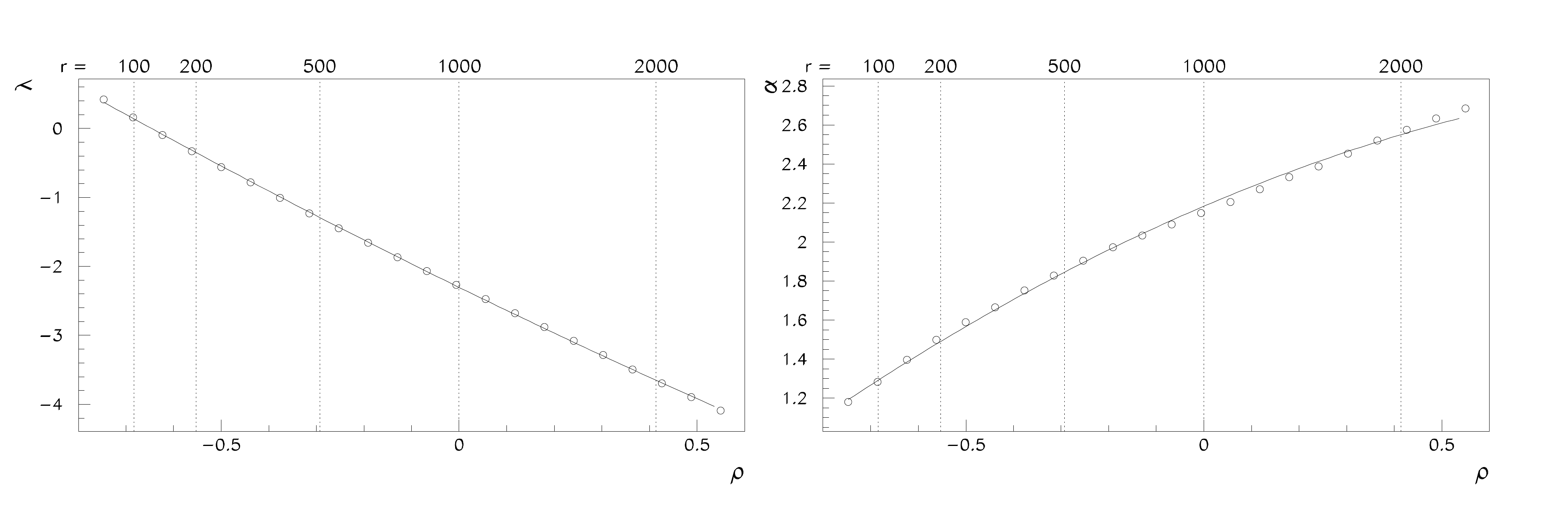,viewport=0 10 1100 340,width=14cm}
\caption{\footnotesize Parametrization of the muon density in ($r,\psi$) coordinates, for a proton shower at 10 EeV, $\theta=80$
  deg, $B_t=60~\mu$T (same organisation as Fig. \ref{fig:map_rphi_1}).}
\label{fig:map_rphi_2}
\end{center}
\end{figure}

\begin{figure}[H]
\begin{center}
\epsfig{file=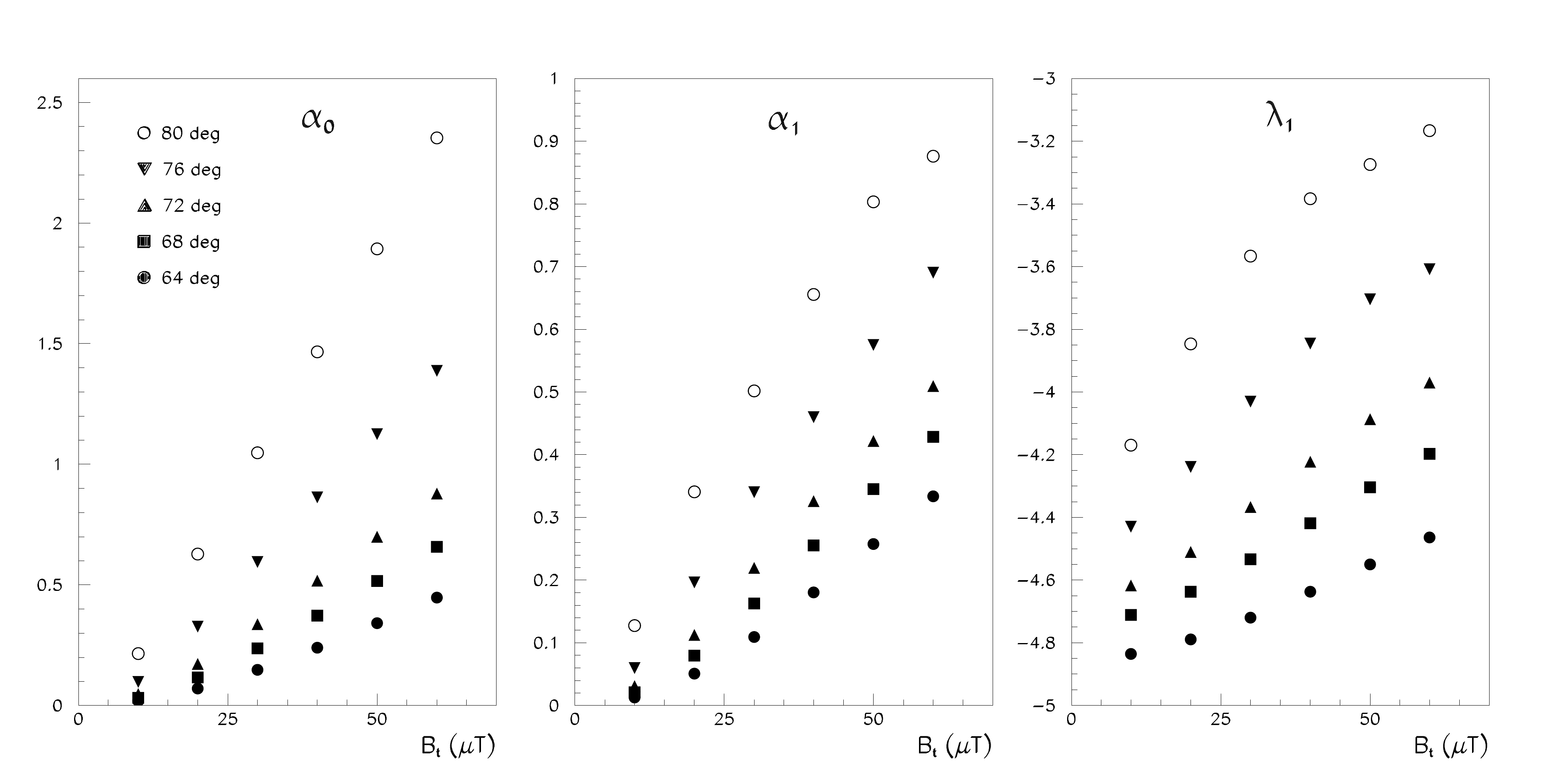,width=15cm}
\caption{\footnotesize Dependence of the parameters of the muon density on the
 zenith angle $\theta$ and the transverse magnetic field $B_t$.}
\label{fig:param_vs_th_bt}
\end{center}
\end{figure}

For each value of $\theta$ and $B_t$ we want to find an analytical expression of the density in the front plane at ground level, as a function
of the distance $r$ to axis and the azimuthal angle $\psi$, with $\psi=0$ in the direction perpendicular to $\overrightarrow{B_t}$. 
Figure \ref{fig:map_rphi_1} and \ref{fig:map_rphi_2} show that for a
wide domain in the useful ranges of $\theta$ and $B_t$ the dependence of the logarithm of
density is approximately linear in $\sqrt{r}$ and sinusoidal in 2$\psi$, with a relative amplitude varying smoothly with $r$.  
So, introducing a reference distance $r_\mathrm{ref}$ (typically the average distance where the density may be measured) 
and the variable $\rho=\sqrt{r/r_\mathrm{ref}}-1$, the density is well fitted by:  
\begin{equation}
f(r,\psi) = \exp\left(\lambda(\rho) +\alpha(\rho) \cos(2(\psi-\psi_B)) \right)~~~~(\psi_B \mathrm{:~direction~of~the~deviation})
\end{equation}

and the dependence on $\rho$ is well decribed by a parabolic parametrization:
\begin{equation}
\lambda(\rho)=\lambda_0+\lambda_1\rho+\lambda_2\rho^2 ~~,~~\alpha(\rho)=\alpha_0+\alpha_1\rho+\alpha_2\rho^2
\end{equation}
$\lambda_0$ is a size parameter, roughly proportional to the primary energy,  while the {\em shape parameters}
$\lambda_1$, $\alpha_0$, $\alpha_1$ carry most of the information on the shape of the ground density that can be extracted
from the measurements;  $\alpha_2$ and $\lambda_2$ are relatively small and need a wide measurement range in $r$ to provide a useful
information. The values of $\lambda_1$, $\alpha_0$, $\alpha_1$ are displayed in Fig. \ref{fig:param_vs_th_bt} as functions of $\theta$
and $B_t$ for an average over 10 proton showers of 1 EeV. 

\subsection{Comparison of the analytical density to the stepwise one}\label{compar_step_anal}
  
\begin{figure}[H]
\begin{center}
\epsfig{file=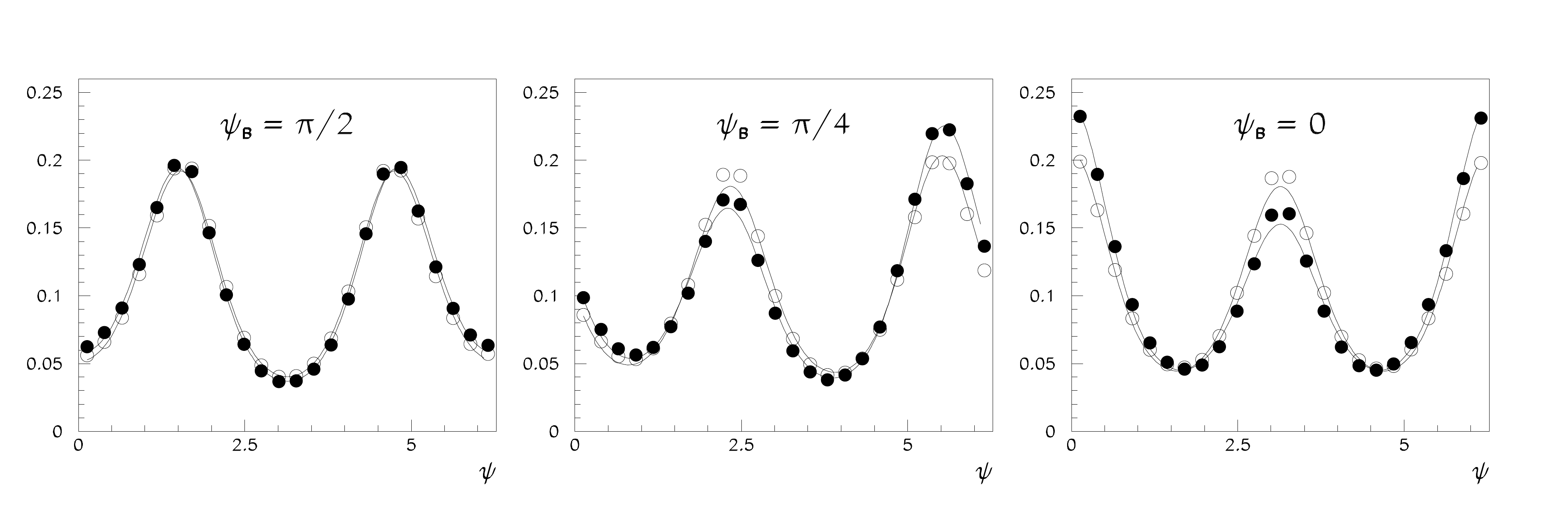,width=15cm}
\caption{\footnotesize Density of muons at $r=1000$ m for a shower of
  1 EeV, at $\theta=72$ deg, with $B_t=60~\mu$T, obtained from the stepwise approximation,
  for different orientations of the transverse magnetic field. Solid: density on ground; open: intrinsic density (projected onto the
 front plane)}
\label{fig:asym_vs_psibt}
\end{center}
\end{figure}

By construction the analytic density has no forward-backward asymmetry in the shower frame, while the stepwise evaluation includes two sources of
asymmetry: the non uniform density of air and the fact that the ground is not perpendicular to the shower axis. Moreover the distortion depends
not only on $|B_t|$, but also on its direction $\psi_B$ in the front plane.  To account for the asymmetry in the stepwise density, 
we introduce in the parametrization a term in $\cos\psi$, that is:
\begin{equation}
f(r,\psi) = \exp\big(\lambda(\rho) +\alpha(\rho) \cos(2(\psi-\psi_B)) +\beta(\rho) \cos(\psi)\big)
\end{equation}
In each slice in $r$ we fit $\lambda$, $\alpha$ and $\beta$ with
this function of $\psi$. An example is given in Fig. \ref{fig:asym_vs_psibt}, which shows that the asymmetry is almost
suppressed when going from the ground densities to the {\em intrinsic} ones, as defined in Sect. \ref{step_extrap}, that is, the ground asymmetry is
dominated by the divergence of the muons from the shower axis. For both options we can extract a parabolic parametrization of $\lambda$ and $\alpha$ as
functions of $\rho$, that is coefficients $\lambda_0,\lambda_1,\lambda_2$ and $\alpha_0,\alpha_1,\alpha_2$,
which are quite similar to the ones obtained with the analytic method.

\section{Correlation of the shape parameters with the nature of the primary}\label{correl_prim}

\subsection{Dependence on shower evolution}\label{dep_evolution}

\begin{figure}[H]
\begin{center}
\epsfig{file=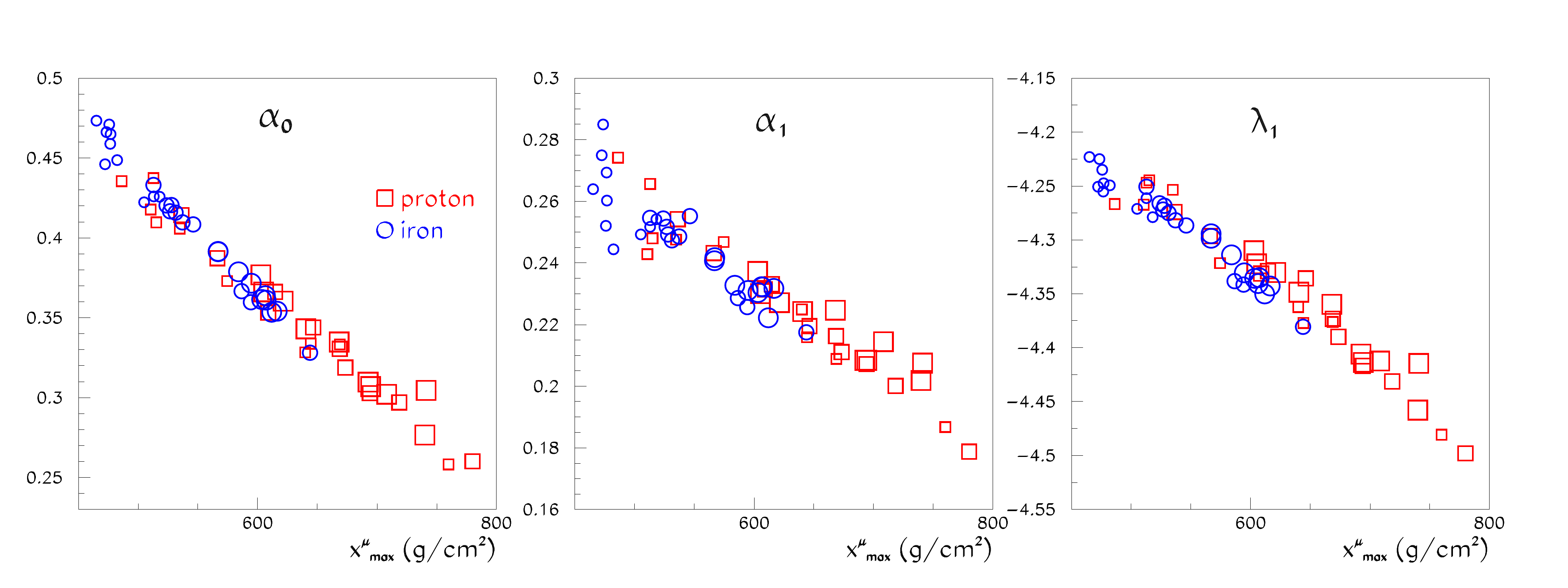,width=15cm}
\caption{\footnotesize Dependence of the shape parameters on $X^\mu_{\rm max}$, for proton and iron showers at different energies, $\theta=72$ deg,
for the QGSJET II-04 model. The size of the symbols is related to the energy: small for 0.1 EeV, medium for 1 EeV, large for 10 EeV.}
\label{fig:param_vs_xmax}
\end{center}
\end{figure}

Here we choose QGSJET II.04 \cite{QGSJET2} as a reference model for the
hadronic interactions. With $\theta=72$ deg, $B_t=$ 30 $\mu$T and $E_{prim}=$ 0.1, 1 and
10 EeV (10 proton and 10 iron showers at each energy), Fig. \ref{fig:param_vs_xmax} shows a tight correlation between the
relevant shape parameters ($\alpha_0$, $\alpha_1$, $\lambda_1$) and $X^\mu_{\rm max}$, the depth of maximum muon production. It is important
to note that proton and iron showers at different energies are on the same line, that is, the shape parameters provide an indirect
measurement  of $X^\mu_{\rm max}$ that can be used for the identification of the primary.
 The same behaviour is observed for the other values of the zenith angle and the transverse field. 
\par Of course the effective identification power (for individual events, or statistically for a sample of events) depends on the
precision that may be achieved on the measurement of the zenith angle
and the shape parameters. This will be discussed in Sect. \ref{precis_meas}. 

\par
\begin{figure}[H]
\begin{center}
\epsfig{file=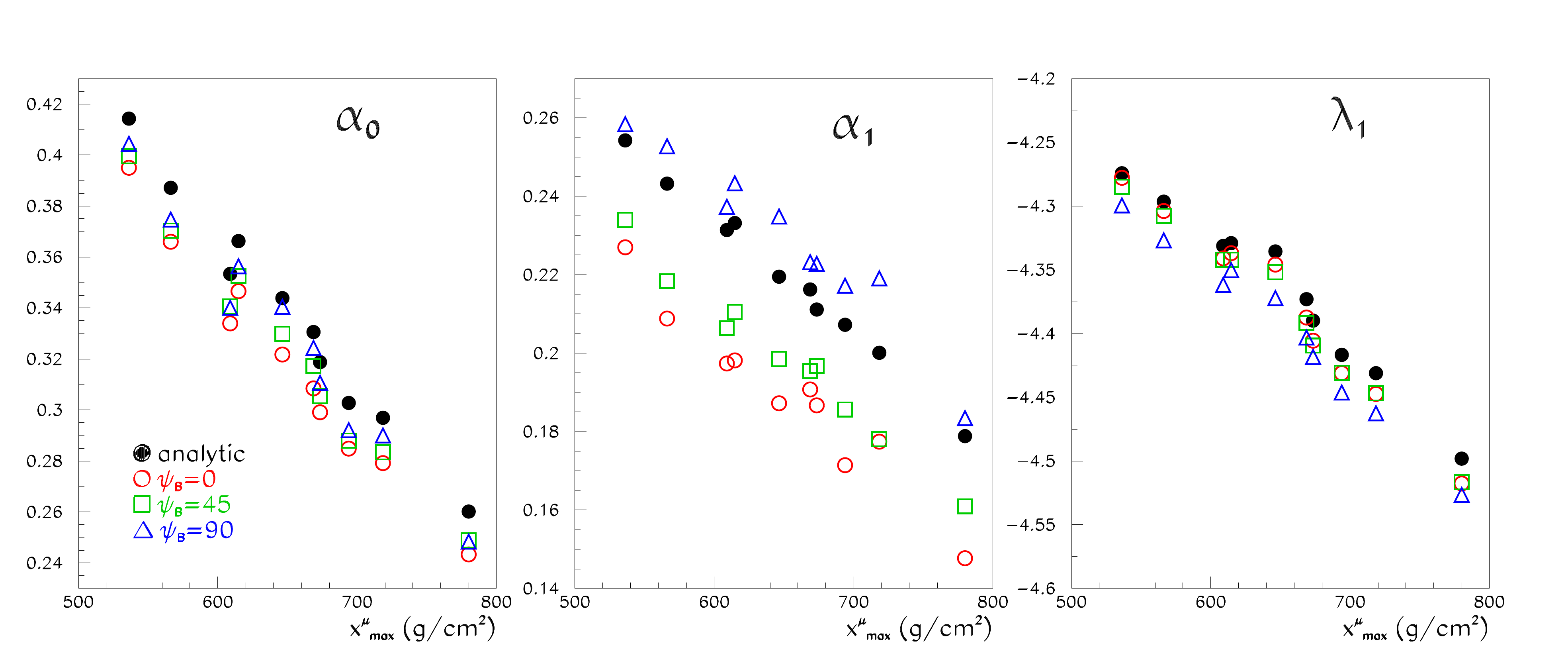,width=15cm}
\epsfig{file=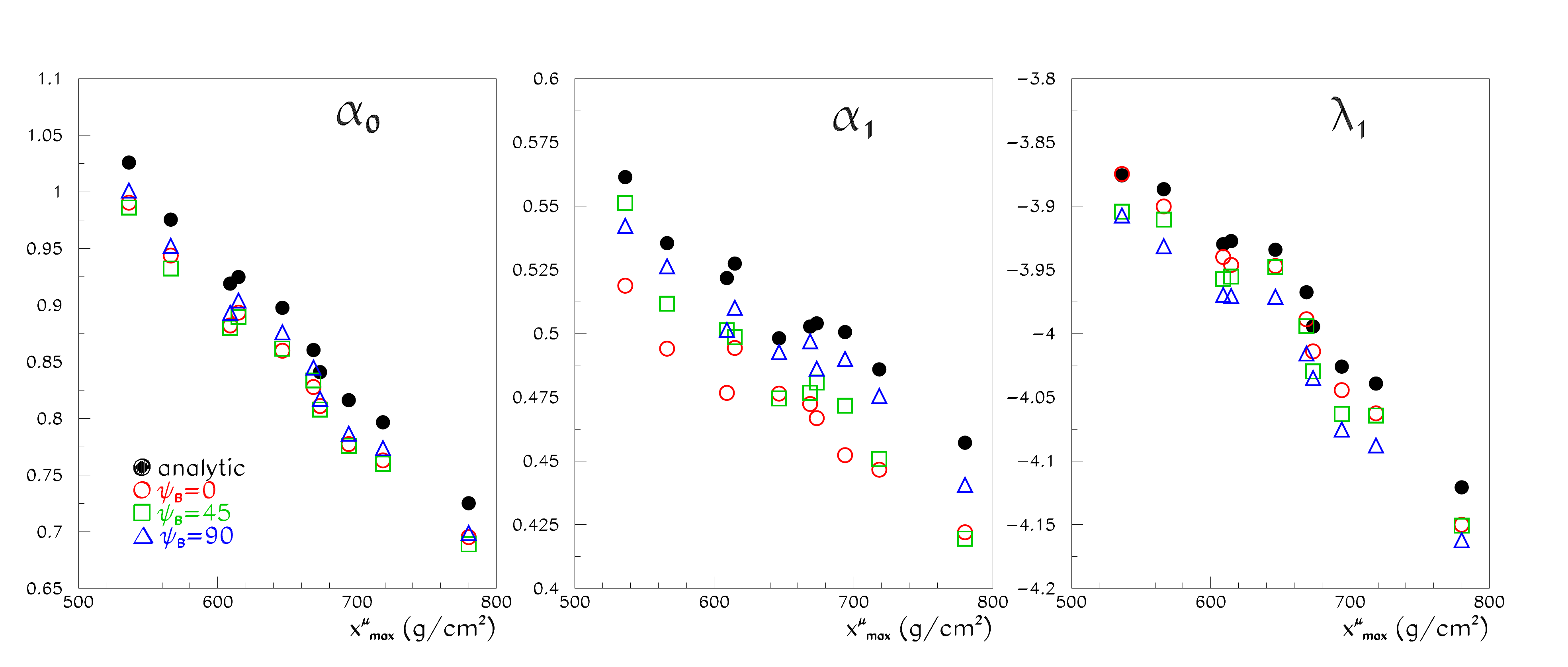,width=15cm}
\caption{\footnotesize Depencence of the shape parameters $\alpha_0$, $\alpha_1$, $\lambda_1$ on $X^\mu_{\rm max}$, obtained
 by the stepwise extrapolation for different orientations of the transverse field (open symbols), and by the analytic method
(solid poins), applied to a proton shower of 1 EeV, at 72 deg. Top: $B_t=30~\mu$T; bottom: $B_t=60~\mu$T.}
\label{fig:compar_step_anal}
\end{center}
\end{figure}

 In principle the density computed with the stepwise extrapolation is more realistic than the analytic
one. Fig. \ref{fig:compar_step_anal} shows that the shape parameters of the density at ground level have the same dependence on
$X^\mu_{\rm max}$ as found with the analytic method. The difference is essentially a global shift, depending on the direction of the field
in the front plane. So we did not apply the stepwise method to the whole set of showers, and we draw conclusions from the results 
of the analytic method.

\subsection{Dependence on hadronic model}\label{dep_model}

\begin{figure}[H]
\begin{center}
\epsfig{file=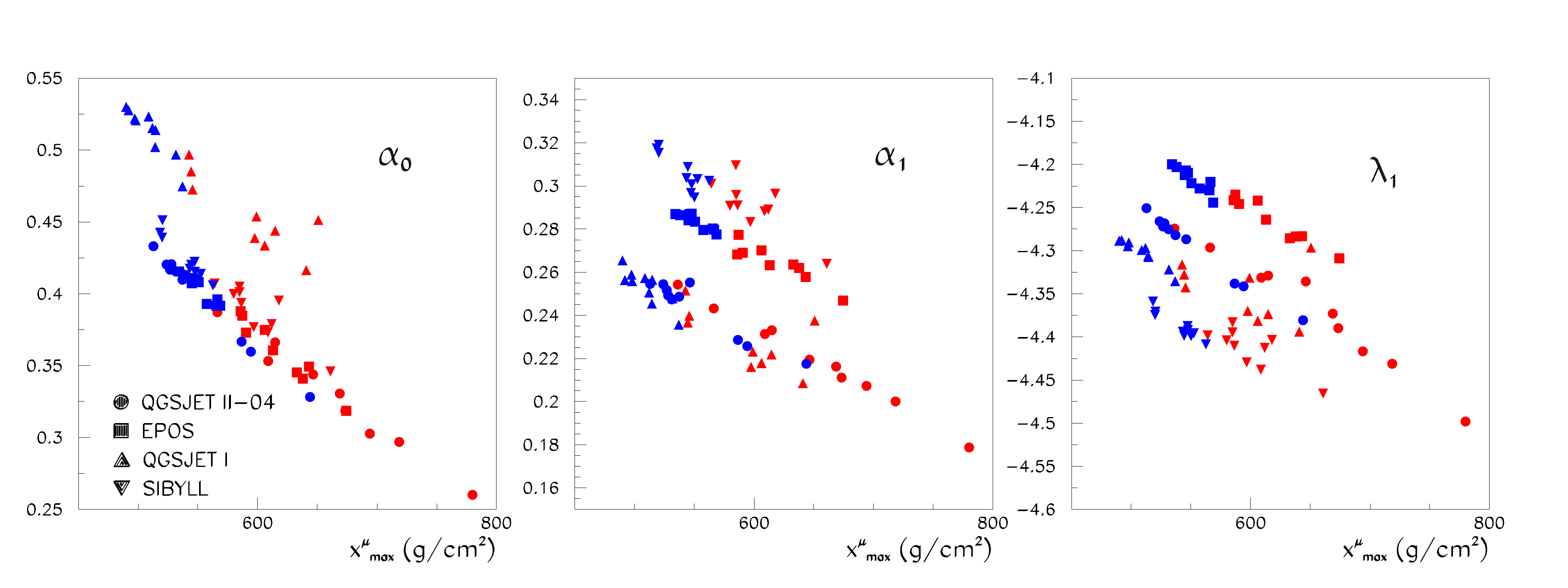,width=15cm}
\caption{\footnotesize Dependence of the shape parameters on $X^\mu_{\rm max}$ at $E=1$ EeV, $\theta=72$ deg, $B_t=30$ $\mu$T,
for different hadronic models. Blue symbols are for iron, red ones for proton.}
\label{fig:compar_param_vs_xmax}
\end{center}
\end{figure}

We have chosen 3 available alternative options: EPOS-LHC \cite{EPOS}, QGSJET1 \cite{QGSJET1} and SIBYLL \cite{SIBYLL}, the last two being
generally considered as obsolete. For each one, we have simulated 10 proton and iron showers at 1 EeV, 72 deg. 
Fig. \ref{fig:compar_param_vs_xmax} shows that for each model the shape parameters have a similar correlation to $X^\mu_{\rm max}$, but the lines
differ significantly for at least one of the parameters. In some cases, the distribution of one parameter may in principle discriminate two models,
whatever the composition. For example, the distribution of $\alpha_1$ separates well QGSJET1 and QGSJET II.04 on the one side from EPOS and
SIBYLL on the other side. Discriminating variables may be defined as combinations of $\alpha_0$, $\alpha_1$, $\lambda_1$.
Again the effective discrimination power depends on the precision of the measurement of these parameters (see Sect. \ref{precis_meas}).

\begin{figure}[H]
\begin{center}
\epsfig{file=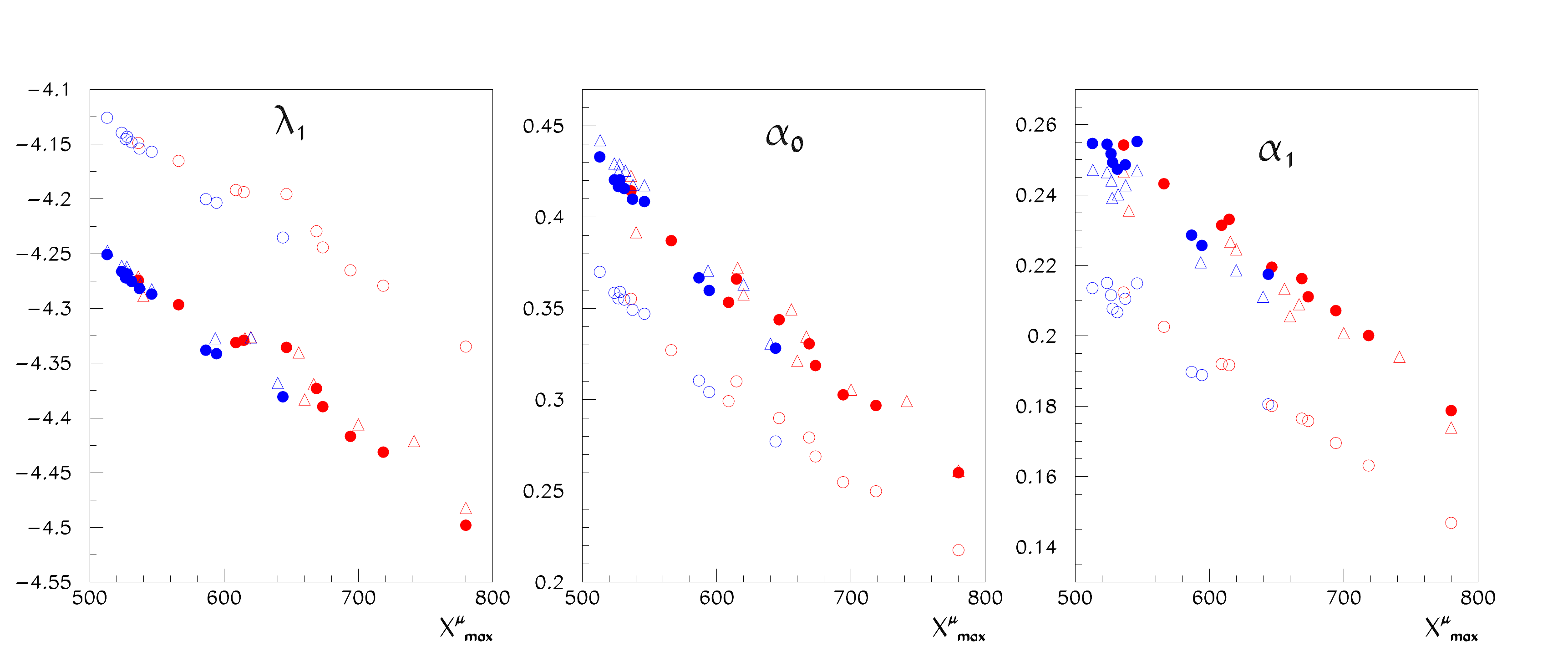,width=15cm}
\caption{\footnotesize Dependence of the shape parameters on $X^\mu_{\rm max}$ at $E=1$ EeV, $\theta=72$ deg, $B_t=30$ $\mu$T, with QGSJET II.04, on artificial modifications. Solid: no modification; open circles: transverse momentum of muons multiplied by 1.1; open triangles: longitudinal profile contracted by 0.9 (see text).}
\label{fig:dep_pt_prof}
\end{center}
\end{figure}

We have also tried to modify by hand some characteristics of the muons. First, by multiplying all transverse momenta by 1.1, because the distribution in $p_t$ may affect the interpretation of the muonic flux at a fixed distance from core. We have also simulated a global contraction of the longitudinal profile of muon production, without modifying $X^\mu_{\rm max}$, by replacing for each muon the depth of production $X_i$ by $X^\mu_{\rm max}+0.9*(X_i- X^\mu_{\rm max)}$, and dividing its energy by the ratio of the air density at the new position to the original one. The results are shown in Fig. \ref{fig:dep_pt_prof}: the contraction of the profile has no significant effect, while the scaling of $p_t$ results in an important shift of the parameters. This means that the magnetic distortion is very sensitive to the distribution of the transverse momentum, which is governed by the latest hadronic interactions.

\section{Precision on the measurements}\label{precis_meas}

We want here to obtain an evaluation of the measurement errors using an ideal detector which is a regular array of identical elements
acting as pure muon counters, that is, giving the number of muons crossing their area $A$ (projected onto the front plane of the shower
axis) with a poissonian distribution. Actually the detector may see the electrons of the electromagnetic halo, and also some photons if
they interact with the material: from this point of view, this approximation is conservative.
\par For the following evaluations we define a {\em standard detector array}: a square array of spacing $\ell=500$ m on a horizontal
ground, with $A=10$ m$^2$. Actually the precision depends essentially on the density of detectors in projection onto the front plane, that is
$1/(\ell^2\cos\theta)$ for the standard array.

\subsection{Precision on zenith angle and transverse field}

\par As can be seen in Fig. \ref{fig:param_vs_th_bt}, the shape parameters depend strongly on the zenith angle $\theta$: if we want to make an
event-by-event  study, we have to be sensitive to a variation of $X^\mu_{\rm max}$ of the order of 30 g/cm$^2$, so we need to
measure $\theta$ with a sufficient precision: for example, around 72
deg, with $B_t=30~\mu$T, the variation of $\alpha_0$ is of the order of 0.04 per degree, which correspond to a variation of 60 g/cm$^2$, so
a precision better than 0.5 degree fulfills the requirement. At higher values of $\theta$ and $B_t$, the dependence in $\theta$ is much
stronger; this is partially compensated by the fact that the dependence of $\alpha_0$ on $X^\mu_{\rm max}$ is also increased,
but not fully. Fortunately the shower front becomes thinner and flatter with increasing $\theta$, so the error on the direction decreases.   
\par The precision on $\theta$ may be evaluated for the {\em standard array} through a conservative approach:
\begin{itemize}
\item making a precise evaluation of $X^\mu_{\rm max}$ for a single event  requires to have measurements on a wide area, at least up to 1000 m from shower axis.
\item  we assume that the precision on the front time measured in a detector is the dispersion of the arrival times of the muons, instead of accounting for the time of the first one, which has less fluctuation in case of multiples hits.
\item we ignore the information from detectors at more than 1000 m from the core  and we attribute to the measurements the dispersion $\sigma_t$ found at 1000 m by the stepwise propagation of muons. This is an overestimation because the dispersion increases with the distance. We find  around 100 ns for $\theta=64$ deg, 50 ns for $\theta=72$ deg and 30 ns for $\theta=80$ deg. We can suppose that the timing precision of the detector is better than 10 ns.
\end{itemize}
With these values, we can evaluate the precision on $\theta$ using the projection onto the front plane. For example, assuming the core to be at the center of one square of the array, with $x'$ axis along the forward-backward direction, we keep $\sim 1/(\ell/1000)^2\cos\theta$ detectors with transverse coordinates $x'_i = \pm(\ell/2)\cos\theta$, $\pm(3\ell/2)\cos\theta$, etc, and we obtain $\sigma_\theta$ through a summation over them:
\begin{equation}
\frac{1}{\sigma^2_\theta} =
\sum_{|x'_i|<1000} \left(\frac{x'_i}{c\sigma_t}\right)^2 \simeq \frac{2}{3} \left(\frac{\ell}{c\sigma_t\cos\theta}\right)^2 
~~~\rightarrow~~~\sigma_\theta \simeq 0.4\sqrt{\cos\theta} \frac{c\sigma_t}{1000}
\end{equation}
(this error scales as $1/\ell$).
With the above values for $\sigma_t$, we find for example es
$\sigma_\theta \simeq 0.5$ deg at $\theta=64$ deg, 0.2 deg at
$\theta=72$ deg and  0.1 deg at $\theta=80$. This is sufficient for our purpose.
\par If a sample of $n$ events is used statistically, what matters is $\sigma_\theta/\sqrt{n}$, so the requirement of the footprint on the
ground may be relaxed. In any case systematic errors, for example the bias due to the ground asymmetry, should kept be under control.

\subsection{Precision on the shape parameters}
  
We suppose that the flux seen by the detectors may be computed in the front plane $(r,\psi)$
with the approximation of Sect. \ref{parametrization}, so that the parametrization found there may be used for the average number of
muons in a detector.\\ 
\par Due to the spacing of the detectors and the rapid decrease at large distance, the range in $r$ providing effective information is
limited; If we choose for $r_\mathrm{ref}$ the medium value of $\sqrt{r}$, we can omit the
terms in $\alpha_2$ and $\lambda_2$ and write the expected number of muons in a detector at position $(r,\psi)$ as:
\begin{equation}
  \overline{N}(r,\psi) = \exp\left(\lambda_0+\lambda_1\rho+(\alpha_0+\alpha_1\rho)\cos(2(\psi-\psi_B))\right)~~~
\mathrm{with}~~\rho=\sqrt{\frac{r}{r_\mathrm{ref}}}-1
\end{equation}

The number of muons in a detector at position $(r_i,\psi_i)$ follows a Poisson law of mean value $\overline{N}(r_i,\psi_i;\mathbf{p})$
depending on the parameters $\mathbf{p}=(x_c,y_c,\lambda_0,\lambda_1,\alpha_0,\alpha_1)$, where $x_c,y_c$ is the core position. 
For given values of $\theta$ and $B_t$ we can also apply this formalism to a ``compound'' parametrization where
$\lambda_1$, $\alpha_0$ and $\alpha_1$ are expressed as linear functions of $X^\mu_{\rm max}$, according the dependence observed
in Sect. \ref{dep_evolution} (Fig. \ref{fig:param_vs_xmax}), with slopes $\lambda'_1=\partial\lambda_1/\partial X^\mu_{\rm max}$,
 $\alpha'_0=\partial\alpha_0/\partial X^\mu_{\rm max}$, $\alpha'_1=\partial\alpha_1/\partial X^\mu_{\rm max}$. In this formalism we have only 
four adjustable parameters $x_c,y_c,\lambda_0,X^\mu_{\rm max}$, and we can obtain directly an uncertainty on $X^\mu_{\rm max}$
according to a given hadronic model. In the following computations, we use as derivative for the fourth parameter:
\begin{equation}
\frac{\partial\overline{N}}{\partial X^\mu_{\rm max}} =  \lambda'_1\frac{\partial\overline{N}}{\partial\lambda_1} 
 + \alpha'_0\frac{\partial\overline{N}}{\partial\alpha_0} + \alpha'_1\frac{\partial\overline{N}}{\partial\alpha_1} 
\end{equation}

\par For a set of measurements the logarithm of the likelihood may be written, omitting the constant terms, as:
\begin{equation}
{\cal L} (\mathbf{p}) = \sum_i \left( n_i\ln(\overline{N}(r_i,\psi_i;\mathbf{p}))  -\overline{N}(r_i,\psi_i;\mathbf{p}) \right)
\end{equation}
Maximizing the likelihood gives an estimator of $\mathbf{p}$ with an error (covariance) matrix $C=W^{-1}$ where the weight matrix $W$ is
defined as:
\begin{equation}
 W_{jk} = -\frac{\partial^2{\cal L}}{\partial p_j \partial p_k} =
 \sum_i \left( \frac{n_i}{\overline{N}_i^2}\frac{\partial\overline{N}_i}{\partial p_j}\frac{\partial\overline{N}_i}{\partial p_k}
+(1-\frac{n_i}{\overline{N}_i}) \frac{\partial^2\overline{N}_i}{\partial p_j \partial p_k} \right)
\end{equation}
where the index $i$ for $\overline{N}$ and its derivatives means ``at position $r_i,\psi_i$''.
In average $n_i=\overline{N}_i$ so we obtain a simple average expression for the elements of $W$:
 \begin{equation}\label{wmatrix}
\overline{W}_{jk} =  \sum_i \frac{1}{\overline{N}_i} \frac{\partial\overline{N}_i}{\partial p_j}\,\frac{\partial\overline{N}_i}{\partial p_k} 
  =  \sum_i \overline{N}_i \frac{\partial(\ln\overline{N}_i)}{\partial p_j}\,\frac{\partial(\ln\overline{N}_i)}{\partial p_k} 
\end{equation}
Various simulations have shown that the weight matrix does not vary strongly in different realizations, so we take the inverse of
$\overline{W}$ as a good estimator of the error matrix we can obtain in real measurements

A direct consequence of Eq.\ref{wmatrix} is the scaling property of the errors: the derivatives $\partial(\ln\overline{N}_i)/\partial p_j$ depend only on
the position, so if $\overline{N}(r,\psi)$ is multiplied by a global factor $F$, the weight matrix is multiplied by $F$, and the errors are
divided by $\sqrt F$. As a consequence, for a given geometrical configuration ($\theta$, $B_t$ and array of detectors),
the errors are approximately proportional to
$1/\sqrt{E_\mathrm{prim}}$. A scaling with the spacing $\ell$ may
be obtained in the approximation of a dense array, if the summation in Eq.\ref{wmatrix}
may be replaced by an integral: this is reasonable here because $\ln(\overline{N}(r,\psi)$ is a smooth function over the
front plane, except at the origin. As a result, $W$ is proportional to $1/(\ell^2\cos\theta)$, so the errors are proportional  to
$\ell\sqrt{\cos\theta}$, that is to the inverse of the square root of the density of detectors in projection onto the front plane.

\subsection{Precision on the indirect measurement of the depth of maximum}

\begin{figure}[H]
\begin{center}
\epsfig{file=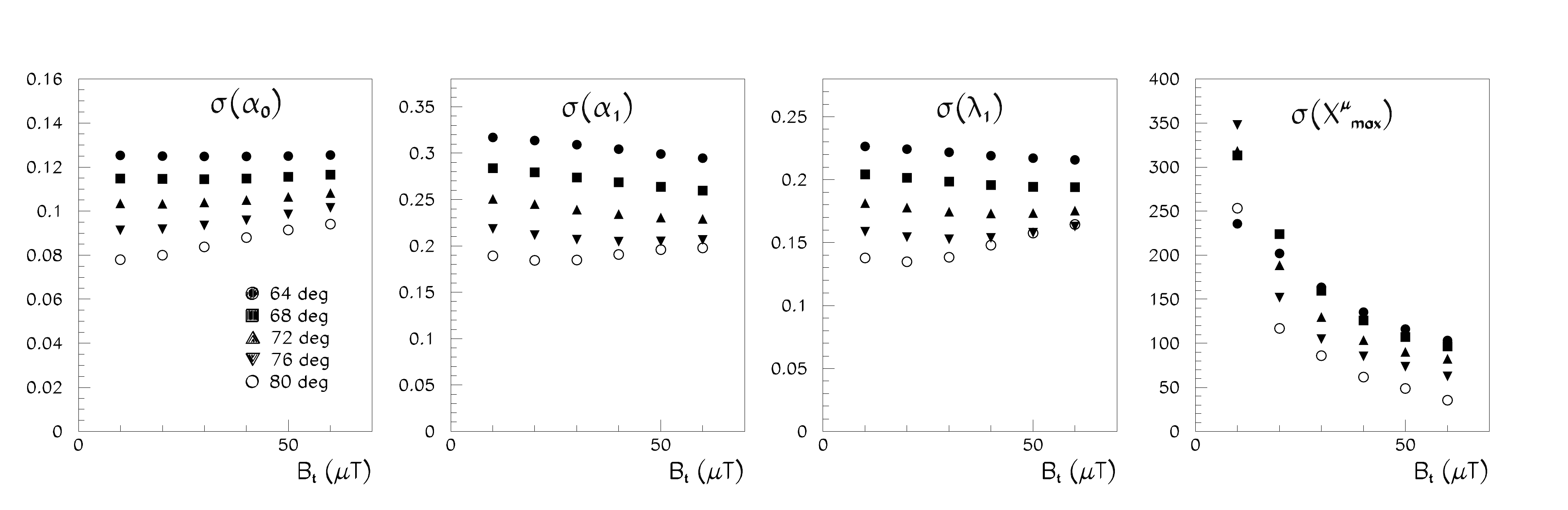,width=15cm}
\epsfig{file=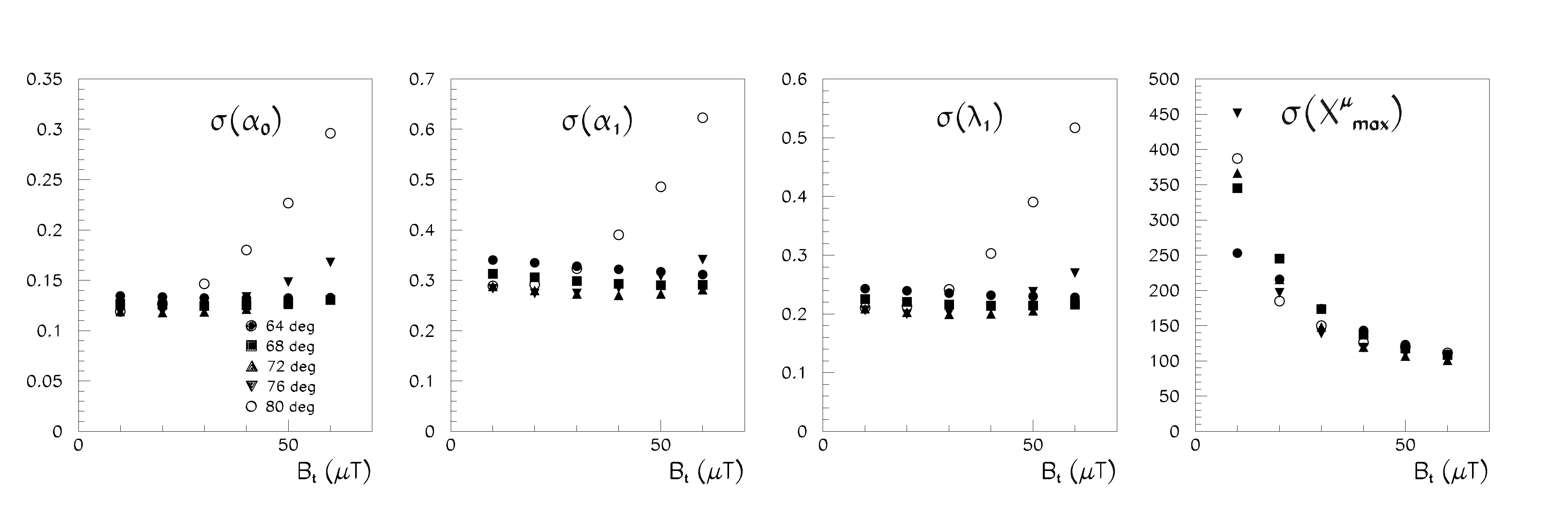,width=15cm}
\caption{\footnotesize Errors on the shape parameters $\alpha_0$, $\alpha_1$ and $\lambda_1$ and induced on $X^\mu_{\rm max}$ as functions of $\theta$ and $B_t$.
 Top: with a normalized size of the muon component (1 hit in average in a detector at 1000 m from core, at 45 degrees from the magnetic field in the transverse
 plane); bottom:  scaled at 1 EeV, with an effective area of 10 m$^2$ per detector).
}
\label{fig:err_xmx}
\end{center}
\end{figure}

Using the values $\lambda_1,\lambda'_1,\alpha_0,\alpha'_0,\alpha_1,\alpha'_1$ obtained
from our sample of simulations of Proton and iron showers at 1 EeV at different values of $\theta$ and $B_t$, we can evaluate using
Eq.\ref{wmatrix} the errors we can expect on an indirect measurement
of $X^\mu_{\rm max}$: the results for the standard array with $r_\mathrm{ref}=1000$ m are plotted on
Fig. \ref{fig:err_xmx}  for $\lambda_0=0$, that is
$\overline{N}(r_\mathrm{ref},\pi/4)=1$ (normalized size of the muon component). The errors on $\alpha_0$,
$\alpha_1$ and $\lambda_1$ depend little on $B_t$ and moderately on $\theta$, while the error on $X^\mu_{\rm max}$ from the compound model
decreases strongly, as expected, with increasing $B_t$. The dependence on $\theta$ is more complicated: for $B_t\geq 30 \mu$T, the precision
on $X^\mu_{\rm max}$ is dominated by the dependence of $\alpha_0$, so the error decreases with increasing $\theta$; at low $\theta$ and
$B_t$, the dependence of $\lambda_1$ is the most important one and may favour a low value of $\theta$. If the errors are computed for a fixed
energy, the dependence on $\theta$ is compensated by the decrease of $\overline{N}$ with $\theta$; using the scaling properties, the result
may be summarized in the following way: for the highest values of $B_t$,  the error on $X^\mu_{\rm max}$ (in g/cm$^2$) is
of the order of $200\ell\sqrt{EA}$ g/cm$^2$, with $\ell$ in km, $E$
in EeV, $A$ in m$^2$.  As a result,  when using a given hadronic model, an event by event identification may be envisaged if the error
on  $X^\mu_{max}$ is about 50 g/cm$^2$ or less, that is, for the standard array, at energies of
the order of 5 EeV or more, if the local geomagnetic field is at least 40 $\mu$T.

\par If the muon detector array is completed by an independent measurement of $X^\mu_{\rm max}$ (or $X_{\rm max}$, which is
tightly correlated), comparing Fig. \ref{fig:compar_param_vs_xmax} and Fig\ref{fig:err_xmx} indicates in which conditions the shape parameters
can provide some discrimination between hadronic models.

\section{Application to possible detectors}\label{detector}

\begin{figure}[H]
\begin{center}
\epsfig{file=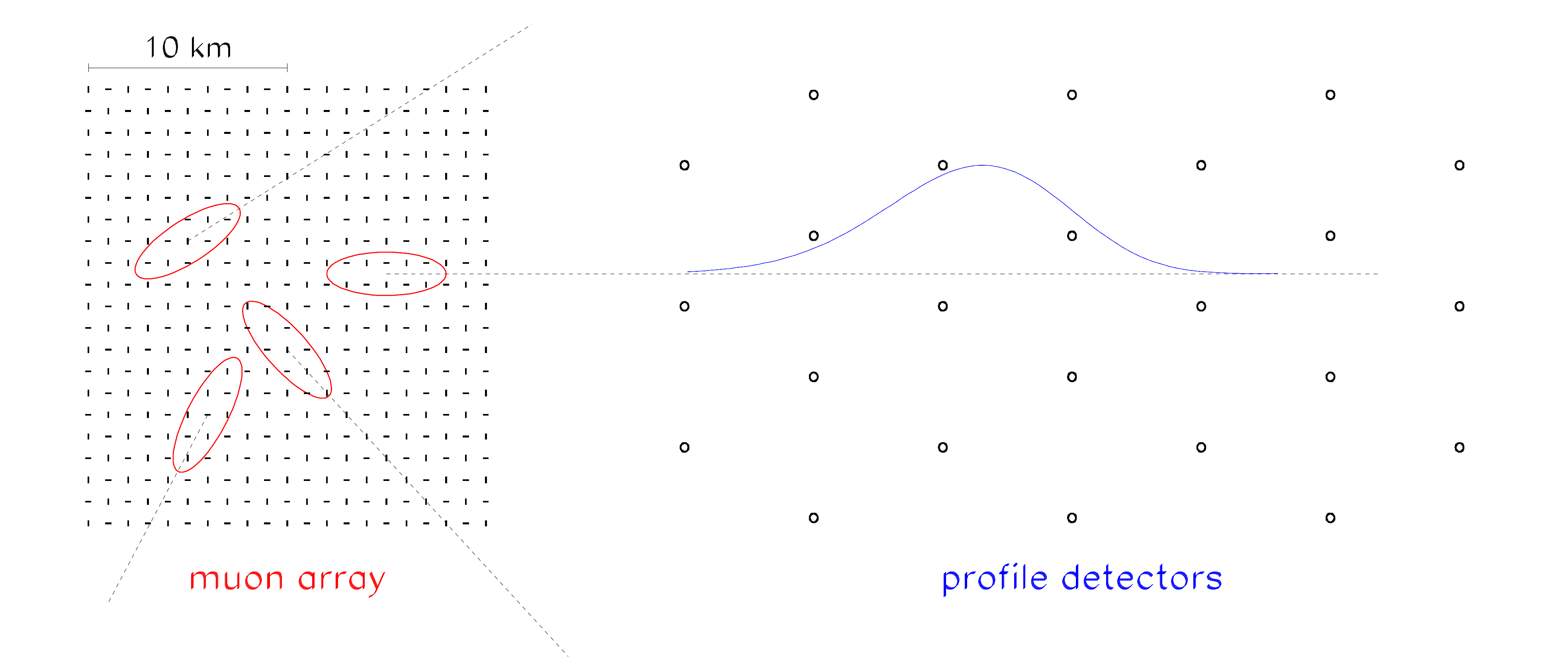,width=15cm}
\caption{\footnotesize A possible layout for a hybrid detector. Left part: muon array: vertical detectors with two orientations to obtain a roughly isotropic sensitivity to horizontal showers; the red ellipses are the contour on ground at 1000 m from the shower axes, for showers at $\theta=70$ deg. Right part: fluorescence eyes or other longitudinal profile detectors; the blue curve is a typical longitufinal profile for such showers.}
\label{fig:array}
\end{center}
\end{figure}

The results found here may be used at different levels of demanding observations, that is also at different levels of dependence on models.
\begin{itemize}
\item Assuming a reliable model of hadronic interaction: an array of muon detectors can obtain an average value of  $X^\mu_{\rm max}$ at lower energies and possibly an event-by-event determination at high energy. As an internal check, the model has to reproduce the observed dependence on $\theta$.
\item Using an external calibration, that is taking the average  $X_{\rm max}$ from another experiment in similar conditions of zenith angle, and correcting for a different atmosphere profile if needed. The hadronic model has to give the right average value of $X^\mu_{\rm max}$: if the model is acceptable, the spectrum of $X^\mu_{\rm max}$ may be exploited, especially on an event-by-event basis at highest energies. This option may be uneasy, because it requires a combination of different experiments. 
\item Building a hybrid detector including a muon array and a longitudinal profile detector covering a few ten kilometers in at least one direction from the muon array, a self-calibrated measurement could be achieved.  The profile detector could be made of single fluorescence eyes with a wide field of view as proposed in \cite{FAST}: for nearly horizontal showers passing above such eyes, the time profile of the received light gives a good measurement of the portion of profile within the field of view. A possible layout of a hybrid detector is proposed in Fig. \ref{fig:array}. Any other detector able to see the profile from the side could replace the fluorescence detection with the advantage of a larger duty cycle.
\end{itemize}

The intensity and the inclination of the magnetic field depends on the location on the surface on the Earth. It has a maximum of about 65
$\mu$T in the North of Asia or in the antarctic region South of Australia, where its direction is nearly vertical,
providing a large value of $B_t$ for nearly horizontal incidence, whatever the azimuth angle.
\par Another important parameter is the altitude. Higher altitude is preferable for higher zenith angle, because the attenuation of the
muon component may prevent an efficient detection after a long path; in the case of a hybrid detector, another advantage is also to reduce the distance between
the ground and the maximum of the longitudinal profile.
\par In this study, we have assumed an horizontal ground, but this is not an experimental requirement. The slope of the ground may increase
the aperture for a restricted azimuthal region, and an elongated shape of the array, with a larger spacing of the detectors in that direction, may be the best option.
\par In any configuration, the profile of the atmosphere needs to be known precisely, because it may be the main source of systematic errors. In case of
large diurnal and/or seasonal variations, it should be carefully monitored.
 
\section{Summary}\label{summary}

Using a semi-analytic calculation of the density of muons in inclined
atmospheric showers, in the approximation of small angular deviations, we have shown that the geomagnetic deflection
 provides an indirect measurement of $X^\mu_{\rm max}$, the depth of maximal production of muons,  
which is an indicator of the nature of the primary particle, tightly correlated to the usual parameter $X_{\rm max}$. This measurement
may be performed using an universal parametrization of the muon density as a function of the distance to the shower axis and the
azimuthal angle in projection onto the front plane (perpendicular to the shower axis at ground level). A significant dependence on $X^\mu_{\rm max}$ was 
found for three parameters $\lambda_1$, $\alpha_1$ and $\alpha_2$, especially the second one which describes the amplitude 
of the quadrupolar distortion of the muon density in the front plane due to the transverse component $\boldmath{B}_t$ of the magnetic field.
The calculation of the density was successfully compared to detailed simulations of the muon propagation in some showers.
\par The parametrization may be inaccurate for very strong distortions, that is for highest values of $\theta$ and/or $B_t$: in such cases the sensitivity to $X_{\rm max}$ is certainly strong and a more accurate simulation of trajectories and a more complex parametrization may be needed to fully exploit the observations. However at large $\theta$ the muonic flux is attenuated and more difficult to measure at ground level; the potential gain on primary identification using events with $\theta > 80$ needs a specific study (maybe with a more elaborated parametrization), beyond the scope of this paper.\\

A precise measurement on the direction is needed as the dependence of the parameters on $\theta$ is important, and increases with $\theta$; fortunately the front of inclined showers is thin, and the thickness decreases with
$\theta$, so a precise measurement of the arrival time of the muons (at the level of 10 ns or better) can determine $\theta$ with enough
precision to extract the dependence of the parameters on $X^\mu_{\rm max}$. 
\par In practice the density is measured on the ground instead of the front plane, so the longitudinal evolution of the shower
and the divergence of muons from the axis result in a forward/backward asymmetry of the observed density: this effect is essentially a dipolar 
distortion  which may be evaluated and corrected for; in any case it may be disentangled from the magnetic distortion. The linear relation between
the distortion parameters and $X^\mu_{\rm max}$ is slightly shifted by a quantity depending on the orientation of the transverse field within
the front plane, but the slope is not modified. A correction for asymmetries of the front at ground level may be needed 
to suppress a possible bias on $\theta$; detailed simulations can estimate such a bias.
\par The variations of the atmospheric density profile at low altitude (below the region of production of the muons) impact mainly the energy loss,
so the usual variations of a few percent do not modify too much the density at ground, and they are easy to monitor.
Variations in upper atmosphere are in principle not large,  but they may affect the development of the hadronic cascade and modify the altitude
of production of the muons and their energy spectrum: this may change both the position of $X^\mu_{\rm max}$ with respect to $X_{\rm max}$,
and the relation between the distortion parameters and $X^\mu_{\rm max}$, which depends on the distance from production to ground. So
for a given site, the profile of the upper atmosphere and its possible variations should be known. \\

We have proposed some possible scenarios of application of the mass sensitive parameters measured with the method developped in this paper. 
Especially, a hybrid detector able to measure simultaneously the longitudinal profile and the magnetic distortion of muons in horizontal showers, can provide strong constraints on the models of hadronic interactions. This option could be valuable to step forward in both the hadronic modelling and in the determination of the composition of UHECRs.

\section*{Acknowledgments}
The work of M.S., made in the ILP LABEX (under reference ANR-10-LABX-63), is supported by French state funds managed by the ANR within the Investissements d'Avenir programme under reference ANR-11-IDEX-0004-02.

\newpage

\section*{Appendix: analytic computation of the ground density with an exponential atmosphere}

To simplify the equations we use the coordinates of the shower frame
($z$ axis along the shower axis), and we omit the $sl$ subscripts. The depth is described by the function:
\begin{equation}
X(z) =  X_a \exp(-z/L)
\end{equation}
where $X_a$ and $L$ account for the slant of the axis.

\par The muons which reach the ground are ultrarelativistic, except possibly in the very end of the trajectory. So we use $E=p$
everywhere\footnote{For convenience we omit $c$ when using the momentum $p$ and the mass $m$, that is, we write $p$ for $pc$ and
$m$ for $mc^2$; however, we keep $c\tau$ in the expression of the decay length.}.
Their direction remains close to the $z$ axis so we can use small angle approximations.

\par Let us consider a muon of mass $m$ and lifetime $\tau$, injected at $z_i$ with energy $E_i$.  When going
down from $z$ to $z-\dd z$, we have to account for:
\begin{itemize}
\item the decay probability: $(-\dd z)/(c\tau E/m)= (-\dd z)/(\lambda E)$ with $\lambda=c\tau/m$.
\item the energy loss: $\dd E=-\varepsilon \dd X= -\varepsilon X_a/L~\exp(-z/L) \dd z$. Introducing the energy extrapolated backwards to infinity
 $E_{\infty} = E_i+\varepsilon X_a/L~\exp(-z_i/L) $, we can write: $E(z)= E_i-\varepsilon \Delta X = E_\infty-\varepsilon  X_a\exp(-z/L)$

\item the magnetic deflection (in the direction perpendicular to the transverse field $\overrightarrow{B_t}$): $\dd \omega = \beta/E(z) \dd
z$ with $\beta=cB_t$ if $E$ is expressed in eV. This elementary deflection produces a deviation $z\dd \omega$ when extrapolated to $z=0$.
\item multiple scattering: a random angular deflection with a variance
  $\dd \eta^2=(E_{ms}/E(z))^2\dd X/X_{rad} =-(E_{ms}/E(z))^2e^{-z/L} \dd z/(LX_{rad})$ in both transverse directions, where $X_{rad}$ is
  the radiation length of air and $E_{ms}=13.6$ MeV.
\end{itemize} 

The cumulative effects at ground level ($z=0$, energy $E_{gr}=E_\infty-\varepsilon X_a$) are then:

\begin{itemize}

\item survival probability $P(z)$:
\begin{eqnarray*}
-\frac{1}{P}\frac{\dd P}{\dd z} &=&  \frac{1}{\lambda\big( E_\infty-\varepsilon X_ae^{-z/L}\big)}~~~\rightarrow~~~~
\dd(\ln P) = -\frac{L}{\lambda  E_\infty} \,\frac{\dd(e^{z/L})}{e^{z/L}-\varepsilon X_a/E_\infty} \\
\ln P(0) &=& -\frac{L}{\lambda E_\infty}\,\ln\left(\frac{E_\infty e^{z_i/L}-\varepsilon X_a}{E_\infty-\varepsilon X_a}\right)
 =  -\frac{L}{\lambda  E_\infty} \, \ln\left(\frac{e^{z_i/L}(E_\infty -\varepsilon X_ae^{-z_i/L})} {E_{gr}}\right) \\
 &=&\frac{L}{\lambda  E_\infty}\ln\left(\frac{X_i}{X_a}\frac{E_{gr}}{E_i}\right)
\end{eqnarray*}

\item angular deviation from $z_i$ to 0:\\
\begin{eqnarray*}
\omega &=& \beta \int_0^{z_i} \frac{\dd z}{E_\infty-X_a\varepsilon
  e^{-z/L}} = \frac{\beta L}{E_\infty}\, \int_0^{z_i} \frac{\dd(e^{z/L})}{e^{z/L}-\varepsilon X_a/E_\infty} \\
&=& \frac{\beta L}{E_\infty} \ln\left(\frac{e^{z_i/L}-\varepsilon X_a/E_\infty}{1-\varepsilon X_a/E_\infty}\right) = 
 \frac{\beta L}{E_\infty} \left(z_i/L+\ln\left(\frac{1-e^{-z_i/L}\varepsilon X_a/E_\infty}{1-\varepsilon X_a/E_\infty}\right)\right) \\
&=& \frac{\beta}{E_\infty} \left( z_i+L\ln\frac{E_i}{E_{gr}}\right) 
\end{eqnarray*} 
\item position deviation from $z_i$ to 0:\\
\begin{eqnarray*}
\delta &=& \beta \int_0^{z_i} \frac{z\dd z}{E_\infty-X_a\varepsilon e^{-z/L}} =\frac{\beta L^2}{E_\infty} F_1(\alpha,z_i/L) 
\end{eqnarray*}
with
\begin{eqnarray*}
\alpha = \frac{\varepsilon X_a}{E_\infty}~~~\mathrm{and}~~~F_1(\alpha,\zeta) = \int_0^\zeta \frac{u\dd u}{1-\alpha e^{-u}}
\end{eqnarray*} 

\item variance of deviation by multiple scattering: we make a
  summation over independent angular deviations in atmosphere slices:
\begin{eqnarray*}
\sigma_{ang}^2 &=&  E_{ms}^2 \int_0^{z_i} \frac{X_ae^{-z/L}\dd z}{LE(z)^2X_{rad}}  =
\left(\frac{E_{ms}}{E_\infty}\right)^2 \frac{X_a}{X_{rad}} \int_0^{z_i}  \frac{-\dd(e^{-z/L}}{(1-\alpha e^{-z/L})^2}  \\
 &=& \left(\frac{E_{ms}}{E_\infty}\right)^2 \frac{X_a}{\alpha X_{rad}} \left(  \frac{1}{1-\alpha}-\frac{1}{1-\alpha e^{-z_i/L}} \right)
   =  \left(\frac{E_{ms}}{E_\infty}\right)^2 \frac{X_a}{\alpha X_{rad}} \frac{E_\infty}{\varepsilon X_a} \left( \frac{E_\infty}{E_{gr}}-\frac{E_\infty}{E_i} \right) \\
   &=&  \frac {E_{ms}^2}{\varepsilon X_{rad}} \left(\frac{1}{E_{gr}}-\frac{1}{E_i}\right)
\end{eqnarray*}
variance of the deviation: we make a summation over the contributions
of the independent angular deviations to the position at $z=0$: 
\begin{eqnarray*}
\sigma_{pos}^2 &=& E_{ms}^2 \int_0^{z_i} \frac{X_ae^{-z/L}z^2\dd z}{LE(z)^2X_{rad}}  =
\left(\frac{E_{ms}}{E_\infty}\right)^2 \frac{X_a}{X_{rad}L} \int_0^{z_i}  \frac{e^{-z/L}z^2\dd z}{(1-\alpha e^{-z/L})^2}  \\
&=&  \left(\frac{E_{ms}}{E_\infty}\right)^2 \frac{X_aL^2}{X_{rad}} F_2(\alpha,\frac{z_i}{L})
\end{eqnarray*}
with 
\begin{eqnarray*}
F_2(\alpha,\zeta) = \int_0^\zeta \frac{e^{-u}u^2\dd u}{(1-\alpha e^{-u})^2}
\end{eqnarray*}

\end{itemize}

\begin{figure}[H]
\begin{center}
\epsfig{file=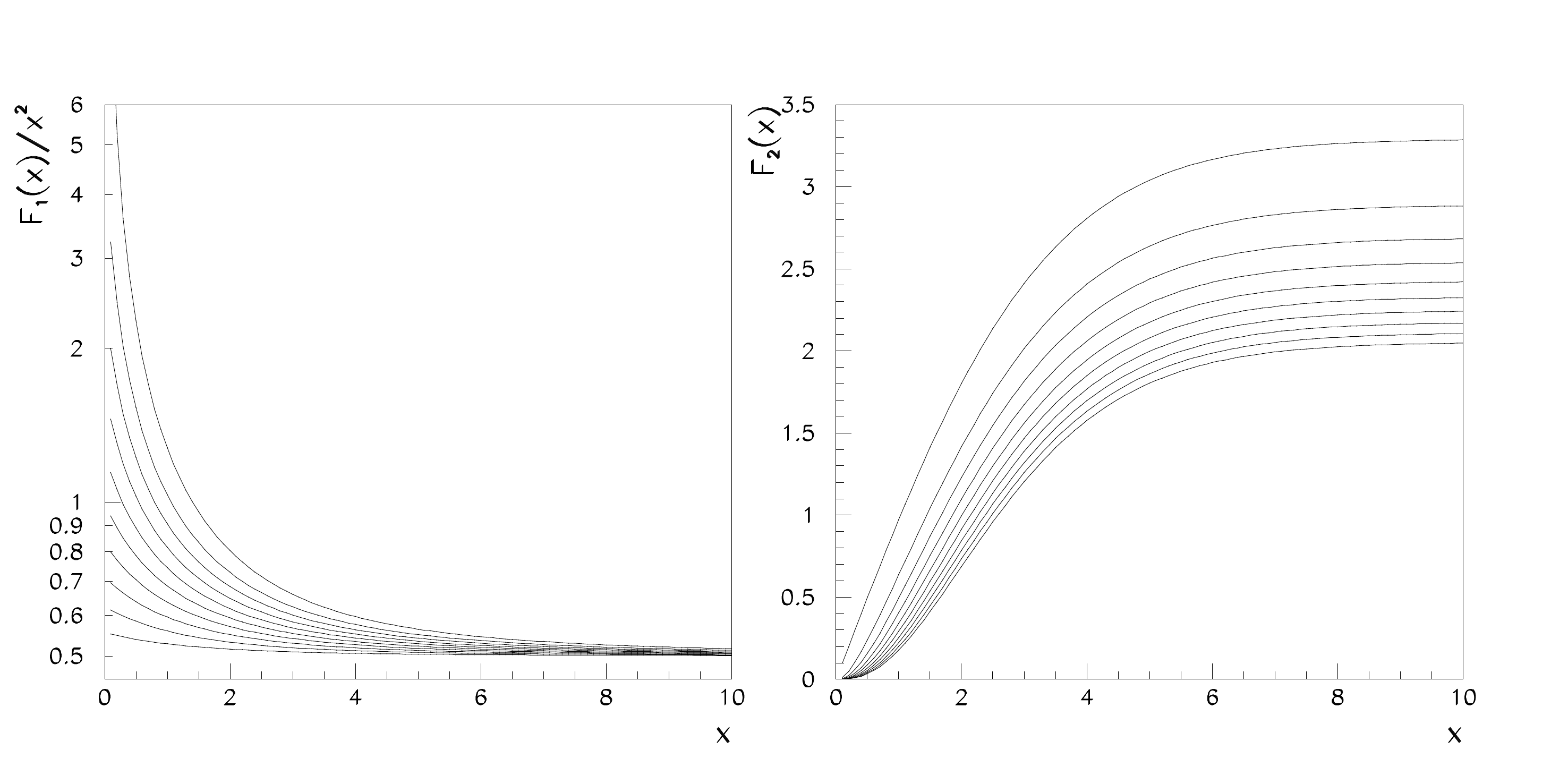,width=16cm}
\caption{\footnotesize Functions $F_1(x)/x^2$ (left) and $F_2(x)$ (right) used to express the magnetic deviation and the multiple scattering dispersion at ground
  level. From bottom to top: $\alpha=0.1, 0.2,\cdots,1$}
\label{fig:tab_func}
\end{center}
\end{figure}

The function $F_2$ may be expressed with $F_1$ and $F_1$ may be
expressed trough the special function Li$_2$ (dilogarithm):
\begin{eqnarray*}
F_2(\alpha,\zeta) &=& \frac{2 F_1(\alpha,\zeta)}{\alpha}-\frac{\zeta^2}{\alpha(1-\alpha e^{-\zeta})} \\
F_1(\alpha,\zeta) &=& \frac{\zeta^2}{2} + \zeta\ln(1-\alpha e^{-\zeta}) + \mathrm{Li}_2(\alpha) - \mathrm{Li}_2(\alpha e^{-\zeta}) \\
\mathrm{Li}_2(x) &=& \sum_{k=1}^{\infty} \frac{x^k}{k^2} = -\int_0^x\frac{\ln(1-u)}{u} \dd u
\end{eqnarray*}
The functions $F_1$ and $F_2$ are plotted in Fig. \ref{fig:tab_func} for different values of $\alpha$ between 0 and 1
(if $\alpha>1$ the muon stops before reaching the ground).

\end{document}